\documentstyle[preprint,eqsecnum,aps,emlines2,bezier]{revtex}
\def\eps{\varepsilon}
\def\partt{\mbox{\boldmath $\partial$}}
\def\const{{\rm const\,}}
\def\gdot{\circle*{20}}

\def\dhline#1#2#3#4#5#6{  {\countdef\nnn=255
\dimendef\llx=0   \dimendef\lly=1
\dimendef\dx=2   \dimendef\dy=3
\llx=#1\unitlength  \lly=#2\unitlength
\dx=#4\unitlength   \dy=#5\unitlength  \nnn=#6
\divide\nnn by 2
\advance\dx by-\llx \advance\dy by-\lly
\div\nnn \div4 \lline \adv
\multiply\dx by2 \multiply\dy by2
\loop \adv \ifnum\nnn>1 \lline \adv \advance\nnn by-1
\repeat \div2 \lline }}

\def\div#1{ \divide\dx by#1  \divide\dy by#1 }
\def\adv{ \advance\llx by\dx \advance\lly by\dy }

\def\lline{ {  \divide\llx by\unitlength \divide\lly by\unitlength
\divide\dx by\unitlength \divide\dy by\unitlength
\advance\dx by\llx \advance\dy by\lly
\emline{\number\llx}{\number\lly}{}{\number\dx}{\number\dy}{}}}

\def\dA{
\begin{picture}(180,140)
\emline{10}{10}{}{90}{130}{}
\emline{90}{130}{}{170}{10}{}
\dhline{20}{25}{}{160}{25}{10}
\emline{38}{39}{}{26}{47}{}
\emline{154}{47}{}{142}{39}{}
\put(90,130){\gdot}
\end{picture}}

\def\dB{
\begin{picture}(180,140)
\emline{10}{10}{}{90}{130}{}
\emline{90}{130}{}{170}{10}{}
\dhline{20}{25}{}{160}{25}{11}
\dhline{50}{70}{}{130}{70}{8}
\emline{36}{36}{}{24}{44}{}
\emline{156}{44}{}{144}{36}{}
\emline{66}{81}{}{54}{89}{}
\emline{126}{89}{}{114}{81}{}
\put(90,130){\gdot}
\end{picture}}

\def\dC{
\begin{picture}(180,140)
\emline{10}{10}{}{90}{130}{}
\emline{90}{130}{}{170}{10}{}
\dhline{20}{25}{}{160}{25}{11}
\emline{34}{33}{}{22}{41}{}
\emline{153}{49}{}{141}{41}{}
\emline{56}{66}{}{44}{74}{}
\emline{82}{105}{}{70}{113}{}
\dhline{40}{55}{}{37}{68}{2}
\dhline{37}{68}{}{38}{81}{2}
\dhline{38}{81}{}{46}{93}{2}
\dhline{46}{93}{}{57}{99}{2}
\dhline{57}{99}{}{70}{100}{2}
\put(90,130){\gdot}
\end{picture}}

\def\dD{
\begin{picture}(180,140)
\emline{10}{10}{}{90}{130}{}
\emline{90}{130}{}{170}{10}{}
\dhline{20}{25}{}{130}{70}{11}
\dhline{50}{70}{}{160}{25}{8}
\emline{38}{39}{}{26}{47}{}
\emline{154}{47}{}{142}{39}{}
\emline{66}{81}{}{54}{89}{}
\emline{126}{89}{}{114}{81}{}
\put(90,130){\gdot}
\end{picture}}

\def\dE{
\begin{picture}(180,140)
\emline{10}{50}{}{90}{130}{}
\emline{90}{10}{}{90}{130}{}
\emline{170}{50}{}{90}{130}{}
\dhline{20}{60}{}{90}{70}{5}
\dhline{90}{25}{}{160}{60}{5}
\emline{83}{40}{}{97}{40}{}
\emline{83}{85}{}{97}{85}{}
\emline{30}{80}{}{40}{70}{}
\emline{150}{80}{}{140}{70}{}
\put(90,130){\gdot}
\end{picture} }

\def\dF{
\begin{picture}(180,140)
\emline{10}{10}{}{90}{70}{}
\emline{10}{130}{}{90}{70}{}
\emline{170}{10}{}{90}{70}{}
\emline{170}{130}{}{90}{70}{}
\dhline{22}{19}{}{22}{121}{8}
\dhline{158}{19}{}{158}{121}{8}
\emline{46}{28}{}{38}{40}{}
\emline{46}{112}{}{38}{100}{}
\emline{134}{28}{}{142}{40}{}
\emline{134}{112}{}{142}{100}{}
\put(90,70){\gdot}
\end{picture} }

\tightenlines

\begin{document}
\draft
\preprint{St Petersburg University Preprint SPbU-IP-98-10;
{\it chao-dyn/9806004}}
\title{Renormalization Group and Anomalous Scaling
in a Simple Model of Passive Scalar Advection in Compressible Flow}
\author{Loran Ts. Adzhemyan and Nikolaj V. Antonov}
\address{ Department of Theoretical Physics, St Petersburg University,
Uljanovskaja 1, St~Petersburg---Petrodvorez, 198904, Russia}
\maketitle
\begin{abstract}
Field theoretical renormalization group methods
are applied to a simple model of a passive scalar quantity
advected by the Gaussian non-solenoidal (``compressible'')
velocity field with the covariance
$\propto\delta(t-t')|{\bf x}-{\bf x'}|^{\eps}$.
Convective range anomalous scaling for the structure functions and
various pair correlators is established, and the corresponding
anomalous exponents are calculated to the order $\eps^{2}$
of the $\eps$ expansion. These exponents are non-universal, as a
result of the degeneracy of the RG fixed point. In contrast to the
case of a purely solenoidal velocity field (Obukhov--Kraichnan model),
the correlation functions in the case at hand exhibit nontrivial
dependence on both the IR and UV characteristic scales, and the
anomalous scaling appears already at the level of the pair correlator.
The powers of the scalar field {\it without derivatives},
whose critical dimensions determine the anomalous exponents,
exhibit multifractal behavior. The exact solution for the pair
correlator is obtained; it is in agreement with the result obtained
within the $\eps$ expansion. The anomalous exponents for passively
advected magnetic fields are also presented in the first order
of the $\eps$ expansion.
\end{abstract}

\section{Introduction}
\label {sec:1}

Much attention has been paid recently to a simple model
of the passive advection of a scalar quantity by a Gaussian
short-correlated velocity field, introduced by Obukhov \cite{Ob}
and Kraichnan \cite{Kraich1}, see the papers [3--24]
and references therein. The structure functions of
the scalar field in this model exhibit anomalous scaling behavior,
and the corresponding anomalous exponents can be calculated
explicitly using certain physically motivated ``linear ansatz''
\cite{Kraich2}, within regular expansions in various small parameters
\cite{Falk1,Falk2,GK,BGK,ShS,Falk3,Pumir,RG},
and using numerical simulations \cite{Kraich3,Lvov,Galanti,VMF}.
On the other hand, this model provides a good testing ground
for various concepts and methods of the turbulence
theory: closure approximations \cite{Kraich2,Kraich3,Yakhot},
refined similarity relations  \cite{Eyink,Xin,Xin2},
Monte Carlo simulations \cite{Yakhot,VMF},
renormalization group \cite{RG}, and so on.

The advection of a passive scalar field $\theta(x)\equiv
\theta(t,{\bf x})$ is described by the stochastic equation
\begin{equation}
\partial _t\theta+ \partial_{i}(v_{i}\theta)
=\nu _0\triangle \theta+f,
\label{1}
\end{equation}
where $\partial _t \equiv \partial /\partial t$,
$\partial _i \equiv \partial /\partial x_{i}$, $\nu _0$
is the molecular diffusivity coefficient, $\triangle$
is the Laplace operator, ${\bf v}(x)$ is the transverse (owing
to the incompressibility) velocity field, and $f\equiv f(x)$ is a
Gaussian scalar noise with zero mean and correlator
\begin{equation}
\langle  f(x)  f(x')\rangle = \delta(t-t')\, C(Mr), \quad
r\equiv|{\bf x}-{\bf x'}|.
\label{2}
\end{equation}
The parameter $L\equiv M^{-1}$ is an integral scale related
to the scalar
noise, and $C(Mr)$ is some function finite as $L\to\infty$.
Without loss of generality, we take $C(0)=1$ (the
dimensional coefficient in  (\ref{2}) can be absorbed by
appropriate rescaling of the field $\theta$ and noise $f$).

In the real problem, the field  ${\bf v}(x)$
satisfies the Navier--Stokes equation.
In the simplified model considered in  [2--8],
${\bf v}(x)$ obeys a
Gaussian distribution with zero average and correlator
\begin{equation}
\langle v_{i}(x) v_{j}(x')\rangle = D_{0}\,
\frac{\delta(t-t')}{(2\pi)^d}
\int d{\bf k}\, P_{ij}({\bf k})\, (k^{2}+m^{2})^{-d/2-\eps/2}\,
\exp [{\rm i}{\bf k}\cdot({\bf x}-{\bf x'})] ,
\label{3}
\end{equation}
where $P_{ij}({\bf k}) = \delta _{ij} - k_i k_j / k^2$ is the
transverse projector, $k\equiv |{\bf k}|$, $D_{0}>0$ is an amplitude
factor, $1/m$ is another integral scale, and $d$ is the
dimensionality of the ${\bf x}$  space; $0<\eps<2$ is a parameter
with the real (``Kolmogorov'') value $\eps=4/3$. The relations
\begin{equation}
D_{0}/\nu_0 \equiv g_{0}\equiv  \Lambda^{\eps}
\label{Lambda}
\end{equation}
define the coupling constant (``charge'') $g_{0}$ and the
characteristic ultraviolet (UV) momentum scale $\Lambda$.

The quantities of interest are, in particular, the single-time
structure functions
\begin{equation}
S_{n}(r)\equiv\langle[\theta(t,{\bf x})-\theta(t,{\bf x'})]^{n}\rangle,
\quad  r\equiv|{\bf x}-{\bf x'}| .
\label{struc}
\end{equation}
In the model (\ref{1})--(\ref{3}), the odd multipoint correlation
functions of the scalar field vanish, while the even single-time
functions satisfy linear partial differential equations \cite{Kraich1},
see also \cite{Falk1,GK,Schlad}.
The solution for the pair correlator is obtained
explicitly; it shows that the structure function $S_{2}$ is
finite for $M$, $m=0$ \cite{Kraich1}. The higher-order
correlators are not found explicitly, but their asymptotic
behavior for $M\to0$ can be extracted from the analysis of
the nontrivial zero modes of the corresponding differential
operators in the limits $1/d\to0$ \cite{Falk1,Falk2},
$\eps\to0$ \cite{GK,BGK}, or $\eps\to2$ \cite{ShS,Pumir}.
It was shown that
the structure functions are finite for $m=0$, and in the
convective range $\Lambda >>1/r>>M$ they have the form
(up to the notation)
\begin{equation}
S_{2n}(r)\propto D_{0}^{-n}\, r^{n(2-\eps)}\, (Mr)^{\Delta_{n}},
\label{HZ1}
\end{equation}
with negative anomalous exponents $\Delta_{n}$, whose first
terms of the expansion in $1/d$ \cite{Falk1,Falk2} and
$\eps$ \cite{GK,BGK} have the form
\begin{equation}
\Delta_{n}=-2n(n-1)\eps/(d+2)+O(\eps^{2})=-2n(n-1)\eps/d +O(1/d^{2}).
\label{HZ3}
\end{equation}

In the paper \cite{RG}, the field theoretical renormalization
group (RG) and operator product expansion (OPE) were applied
to the model (\ref{1})--(\ref{3}). In the RG approach, the
anomalous scaling for the structure functions and
various pair correlators is established as a consequence of
the existence in the corresponding operator product expansions
of ``dangerous'' composite operators [powers of the local
dissipation rate], whose {\it negative} critical dimensions
determine the anomalous exponents $\Delta_{n}$.
The exponent $\eps$ plays in the RG approach the role analogous to
that played by the parameter $\eps=4-d$ in the RG theory of critical
phenomena \cite{Zinn}. The anomalous exponents
were calculated in \cite{RG}
to the order $\eps^{2}$ of the $\eps$ expansion for the arbitrary
value of $d$, and they are in agreement with the first-order results
obtained in the zero-modes approach \cite{Falk1,Falk2,GK,BGK}.
The RG approach to the stochastic theory of turbulence is reviewed
in \cite{UFN}.

In the paper \cite{Kraich1}, a closure-type approximation for
the model (\ref{1})--(\ref{3}), the so-called linear ansatz, was
used to derive simple
explicit expression for the anomalous exponents for any $0<\eps<2$,
$d$, and $n$. Although the predictions of the linear ansatz appear
consistent with some numerical simulations \cite{Kraich2,VMF} and
exact relations \cite{Gat,Kraich4}, they do not agree with the
results obtained within the zero-modes and RG approaches in the
ranges of small  $\eps$, $2-\eps$ or $1/d$.
This disagreement can be related to the fact that these limits
have strongly nonlocal dynamics in the momentum space,
which suggests possible relation
between deviations from the linear ansatz and locality of the
interactions, see discussion in Refs.\cite{Kraich4,VMF}.
\footnote{ The small $\eps$ limit can be treated perturbatively,
the effective small parameter equals to the reciprocal of the
significant range of interactions in the momentum space.
This range becomes infinite as $\eps$ goes to zero \cite{private}.}

The results of the RG approach are completely reliable and
internally consistent for small $\eps$, but the validity of
their extrapolation to the finite values of $\eps$ is not obvious.
Most numerical simulations have been limited to two dimensions
\cite{Kraich3,Galanti}
and have not yet been able to cover the small $\eps$ or large
$d$ ranges, in which the reliable analytical results are available.
Therefore, it is not yet clear whether the anomalous scaling in the
small $\eps$  and finite $\eps$ ranges has the same origin,
with the exponents depending continuously on $\eps$,
or there is a ``crossover'' in the anomalous scaling behavior
for some small but finite value of $\eps$ and these ranges should
be treated separately.

Another important question is that of the universality of
anomalous exponents. The exponents $\Delta_{n}$ in (\ref{HZ3})
do not depend on the choice of the correlator (\ref{2}) and on
the specific form of the infrared (IR) regularization in the
correlator (\ref{3}).
It was argued on phenomenological grounds in \cite{ShS}
that the anomalous exponents in the Gaussian model can depend on
more details of the velocity statistics than the exponent $\eps$.
The exponents indeed change when the function $\delta(t-t')$ in
the correlator (\ref{3}) is replaced by some function
with finite width, i.e., the velocity has short but finite
correlation time \cite{Falk3}, and when the velocity field is taken to
be time-independent (see Sec. V of the Ref.\cite{RG}).

In this paper, we consider the generalization of the model
(\ref{1})--(\ref{3})  to the case of a non-solenoidal
(``compressible'') velocity field. In this case, the correlator
(\ref{3}) is replaced by
\begin{equation}
\langle v_{i}(x) v_{j}(x')\rangle =
\frac{\delta(t-t')}{(2\pi)^d}
\int d{\bf k}\,
\frac{D_{0}\,P_{ij}({\bf k})+D_{0}'\,Q_{ij}({\bf k})}
{(k^{2}+m^{2})^{d/2+\eps/2}}\,
\exp [{\rm i}{\bf k}\cdot({\bf x}-{\bf x'})] .
\label{4}
\end{equation}
The notation is explained below the Eq. (\ref{3}); the new
quantities are the longitudinal projector
$Q_{ij}({\bf k}) =  k_i k_j / k^2$ and the
additional amplitude factor $D_{0}'>0$.

One should not expect that a Gaussian, white-noise model
of the type (\ref{1}), (\ref{2}), (\ref{4}) provides very
good approximation for the real compressible advection;
however, it can be used to illustrate the important distinctions
which exist between the compressible and incompressible cases,
see e.g. \cite{El,Avell} and references therein.

The aim of this paper is to give the RG treatment of anomalous
scaling with non-universal exponents; to compare the results of
the $\eps$ expansion with the nontrivial exact exponent, and to
present analytic results which probably will be easier to compare
with numerical simulations than the analogous results for the
incompressible case. We apply the RG method to the model
(\ref{1}), (\ref{2}), (\ref{4}) to establish the existence of
the anomalous scaling in the convective range and to calculate
the corresponding anomalous exponents to the second order of the
$\eps$ expansion. We show that the single-time two-point
correlation functions of the powers of the scalar field
in the convective range have the form
\begin{equation}
\langle\theta^{n}(t,{\bf x})\theta^{p}(t,{\bf x'})\rangle\propto
\nu_0^{-(n+p)/2}\,\Lambda^{-(n+p)}\,
(\Lambda r)^{-\Delta_{n}-\Delta_{p}} (Mr)^{\Delta_{n+p}},
\quad  r\equiv|{\bf x}-{\bf x'}| .
\label{as1}
\end{equation}
for even $n+p$ and zero otherwise.
In addition to $\eps$  and $d$, the exponents $\Delta_{n}$
depend on a free parameter, the ratio $\alpha\equiv D_{0}'/D_{0}$
of the amplitudes in the correlator (\ref{4}). In the first order
of the expansion in $\eps$ they have the form
\begin{equation}
\Delta_{n}=n(-1+\eps/2)-\frac{\alpha n(n-1)d\eps}{2(d-1+\alpha)}
+O(\eps^{2})
\label{exp}
\end{equation}
(the results $\Delta_{1}=-1+\eps/2$ for any $\alpha$ and
$\Delta_{n}=n(-1+\eps/2)$ for $\alpha=0$ are in fact exact).
We have also calculated the $\eps^{2}$ term of the
exponent $\Delta_{n}$
for any $d$ and $\alpha$; the result is rather cumbersome and
is given in Sec. \ref{sec:3}.

The leading term of the convective range behavior of the
structure functions (\ref{struc}) in the model (\ref{4})  is
completely determined by the contribution $\langle\theta^{2n}\rangle$;
it is obtained from (\ref{as1}) by the substitution $n\to2n$,
$p\to0$ and has the form
\begin{equation}
S_{2n}(r)\propto \nu_0^{-n}\, \Lambda^{-2n}\,
(M/\Lambda)^{\Delta_{2n}}.
\label{as2}
\end{equation}

It follows from (\ref{as1}) that the anomalous scaling
in the model (\ref{4}) appears already at the
level of the pair correlation function. The corresponding exponent
$\Delta_{2}$ is found exactly for all $0<\eps<2$ from the exact
solution for the single-time pair correlator, see Sec. \ref{sec:2}:
\begin{equation}
\Delta_{2}= -2-
\frac{\eps\, (\alpha-1)\, (d-1)}{(d-1)+\alpha\,(1+\eps)}
\label{exp2}
\end{equation}
(anomalous scaling for the pair correlator with the
exactly known exponent was established previously in \cite{MHD} on
the example of a passively advected magnetic field).

In the language of the RG, the non-universality of the exponents
(\ref{exp}), (\ref{exp2}) is explained by the fact that the fixed
point of the RG equations is degenerated: its coordinate depends
continuously on the ratio $\alpha$ (see Sec. \ref{sec:3}).

In contradistinction with the model (\ref{3}), where the anomalous
exponents are related to the critical dimensions of the composite
operators $(\partial_{i}\theta\partial_{i}\theta)^{n}$ \cite{RG},
the exponents $\Delta_{n}$ in (\ref{as1}), (\ref{as2}) are determined by
the critical dimensions  of the monomials $\theta^{n}$, the powers
of the field itself, and these dimensions appear to be nonlinear
functions of $n$, see Sec. \ref{sec:4}. This explains the difference
between the convective range behavior of the model (\ref{3}) and
that of the model (\ref{4}) and makes the limit $D'_{0}\to0$
rather subtle.

The model (\ref{4}) remains nontrivial in the case $d=1$,
where the velocity field becomes purely potential. One can
hope that the one-dimensional case is more accessible
to study using numerical simulations, than the lowest-dimensional case
$d=2$ for (\ref{3}), and it will be possible to compare
the analytic results (\ref{as1})--(\ref{exp2}) with the
numerical estimates (despite the fact that the structure
functions (\ref{as2}) are independent of $r$, the values
of the anomalous exponents can be extracted from their dependence
on $M$).
In the paper \cite{VM}, the model (\ref{4}) has been studied
directly for the one-dimensional case in terms of certain potential
function for the field $\theta$; the analytic expressions for the
anomalous exponents obtained within the zero-modes technique have
been found to agree with non-perturbative numerical results.
The relationship between our results and the results of \cite{VM}
is discussed in Sec. \ref{sec:2}.

The paper is organized as follows. In Sec. \ref{sec:2},
we give the field theoretical formulation of the model
(\ref{1}), (\ref{2}), (\ref{4}) and derive exact
equations for the response function and pair correlator of
the scalar field. The explicit solution for the pair correlator
is obtained and the exact expression (\ref{exp2}) for the
corresponding anomalous exponent is derived.
In Sec. \ref{sec:3}, we perform the UV renormalization of the model and
derive the corresponding RG equations with exactly known RG
functions (the $\beta$ function and the anomalous dimension).
These equations have an IR stable fixed point, which establishes
the existence of IR scaling with exactly known critical
dimensions of the basic fields and parameters of the model.
The solution of the RG equations for the correlation
functions (\ref{as1}) is given, which determines their dependence
on the UV scale. In Sec. \ref{sec:4}, the dependence of the correlators
on the IR scale is studied using the OPE, and the relations (\ref{as1}),
(\ref{as2}) are derived. We also discuss briefly the RG approach to
the model of passively advected magnetic fields introduced in \cite{MHD}.
In Sec. \ref{sec:5}, we present the calculation of the anomalous
exponents in the model (\ref{4}) to the order $\eps^{2}$
of the $\eps$ expansion.
The results obtained are briefly discussed in Sec. \ref{sec:6}.

\section{Exact solution for the pair correlation function}
\label {sec:2}

The single-time correlation functions of the field $\theta$ in the
models of the type (\ref{1}), (\ref{2}), (\ref{3}) or (\ref{4})
satisfy closed linear partial differential equations
\cite{Kraich1} (see also Refs.\cite{Falk1,GK,Schlad}). Below we
give an alternative derivation of the equation for the pair
correlation functions based on the field theoretical formulation
of the problem.

The stochastic problem (\ref{1}), (\ref{2}), (\ref{4}) is equivalent
to the field theoretical model of the set of three fields
$\Phi\equiv\{\theta, \theta',{\bf v}\}$ with action functional
\begin{equation}
S(\Phi)=\theta' D_{\theta}\theta' /2+
\theta' \left[ - \partial_{t}\theta -\partt ({\bf v}\theta)
+ \nu _0\triangle \theta \right]
-{\bf v} D_{v}^{-1} {\bf v}/2.
\label{action}
\end{equation}
The first four terms in  (\ref{action}) represent
the Martin--Siggia--Rose-type action  \cite{MSR,Ja,Ja2,DeDom1}
for the stochastic problem (\ref{1}), (\ref{2}) at fixed ${\bf v}$,
and the last term
represents the Gaussian averaging over ${\bf v}$. Here $D_{\theta}$
and $D_{v}$ are the correlators (\ref{2}) and (\ref{4}), respectively,
the required integrations over $x=(t,{\bf x})$ and summations over
the vector indices are understood.

The formulation (\ref{action}) means that statistical averages
of random quantities in stochastic problem (\ref{1}), (\ref{2}), (\ref{4})
coincide with functional averages with the weight $\exp S(\Phi)$,
so that the generating functionals of total ($G(A)$) and connected
($W(A)$) Green functions of the problem
are represented by the functional integral
\begin{equation}
G(A)=\exp  W(A)=\int {\cal D}\Phi \exp [S(\Phi )+A\Phi ]
\label{field}
\end{equation}
with arbitrary sources $A\equiv A^{\theta},A^{\theta'},A^{\bf v}$
in the linear form
\[A\Phi \equiv \int dx[A^{\theta}(x)\theta (x)+A^{\theta '}(x)\theta '(x)
+ A^{\bf v}_{i}(x)v_{i}(x)].\]

The model (\ref{action}) corresponds to a standard Feynman
diagrammatic technique with the triple vertex
$-\theta'\partt({\bf v}\theta)\equiv\theta' V_{j}v_{j}\theta$
with vertex factor (in the momentum-frequency representation)
\begin{equation}
V_{j}={\rm i} k_{j},
\label{vertex}
\end{equation}
where ${\bf k}$ is the momentum flowing into the vertex via the
field $\theta'$.
The bare propagators in the momentum-frequency representation
have the form
\begin{mathletters}
\label{lines}
\begin{equation}
\langle \theta \theta' \rangle _0=\langle \theta' \theta \rangle _0^*=
(-i\omega +\nu _0k^2)^{-1} ,
\label{lines1}
\end{equation}
\begin{equation}
\langle \theta \theta \rangle _0=C(k)\,(\omega ^2+\nu _0^2k^4)^{-1},
\label{lines2}
\end{equation}
\begin{equation}
\langle \theta '\theta '\rangle _0=0 ,
\label{lines3}
\end{equation}
\end{mathletters}
where $C(k)$ is the Fourier transform of the function $C(Mr)$ from
(\ref{2}) and the bare propagator $\langle{\bf v}{\bf v}\rangle _0$
is given by Eq. (\ref{4}). The parameter $g_{0}\equiv D_{0}/\nu_0$
plays the part of the coupling constant in the perturbation theory.
The pair correlation functions $\langle\Phi\Phi\rangle$ of the
multicomponent field $\Phi$ satisfy standard Dyson equation, which in
the component notation reduces to the system of two equations, cf.\cite{Wyld}
\begin{mathletters}
\label{Dyson}
\begin{equation}
G^{-1}(\omega, k)= -{\rm i}\omega +\nu_0 k^{2} -
\Sigma_{\theta'\theta} (\omega, k),
\label{Dyson1}
\end{equation}
\begin{equation}
D(\omega, k)= |G(\omega, k)|^{2}\, [C(k)+\Sigma_{\theta'\theta'} (\omega, k)],
\label{Dyson2}
\end{equation}
\end{mathletters}
where $G(\omega, k)\equiv\langle\theta\theta'\rangle$ and
$D(\omega, k)\equiv\langle\theta\theta\rangle$ are the exact response
function and pair correlator, respectively, and $\Sigma_{\theta'\theta}$,
$\Sigma_{\theta'\theta'}$ are self-energy operators represented by
the corresponding 1-irreducible diagrams; the functions
$\Sigma_{\theta\theta}$, $\Sigma_{{\bf v}{\bf v}}$ in the model
(\ref{action}) vanish identically.

The feature characteristic of the models like (\ref{action}) is that
all the skeleton multiloop diagrams entering into the
self-energy operators
$\Sigma_{\theta'\theta}$, $\Sigma_{\theta'\theta'}$
contain effectively closed circuits of retarded
propagators $\langle\theta\theta'\rangle$  and therefore
vanish (it is also crucial here that the propagator $\langle
{\bf v}{\bf v}\rangle_{0}$ in (\ref{4}) is proportional to the
$\delta$ function in time). Therefore the self-energy operators
in (\ref{Dyson}) are given by the single-loop approximation
exactly and have the form\footnote{ The single-loop
approximation to the Dyson equations in the stirred
hydrodynamics is equivalent \cite{Wyld} to the well-known direct
interaction approximation (DIA) \cite{DIA}. One can say that in the models
(\ref{1}), (\ref{2}), (\ref{3}) or (\ref{4}) the DIA appears to be exact.}
\begin{mathletters}
\label{sigma}
\begin{equation}
\Sigma_{\theta'\theta} (\omega, k)= -\int\frac{d\omega'}{2\pi}
\int\frac{d{\bf q}}{(2\pi)^{d}}
\frac{D_{0} (k^{2}-({\bf k}\cdot{\bf q})^{2}/q^{2}) +
D_{0}'({\bf k}\cdot{\bf q})^{2}/q^{2}} {(q^{2}+m^{2})^{d/2+\eps/2}}\,
G(q',\omega'),
\label{sigma1}
\end{equation}
\begin{equation}
\Sigma_{\theta'\theta'} (\omega, k)=\int\frac{d\omega'}{2\pi}
\int\frac{d{\bf q}}{(2\pi)^{d}}
\frac{D_{0} (k^{2}-({\bf k}\cdot{\bf q})^{2}/q^{2}) +
D_{0}'({\bf k}\cdot{\bf q})^{2}/q^{2}} {(q^{2}+m^{2})^{d/2+\eps/2}}\,
D(q',\omega'),
\label{sigma2}
\end{equation}
\end{mathletters}
where $q'\equiv|{\bf k}-{\bf q}|$.
The integrations over $\omega'$ in the right hand sides of
Eqs. (\ref{sigma}) give the single-time
response function $G(q)=(1/{2\pi})\int{d\omega'}\,G(q,\omega')$
and the single-time pair correlator
$D(q)=(1/{2\pi})\int{d\omega'}\,D(q,\omega')$
(note that the both self-energy operators are in fact independent of
$\omega$). The only contribution to $G(q)$ comes from the bare
propagator (\ref{lines1}), which in the $t$ representation is
discontinuous at coincident times. Since the correlator (\ref{4}),
which enters into the single-loop diagram for $\Sigma_{\theta'\theta}$,
is symmetric in $t$ and $t'$, the response function must be
defined at $t=t'$ by half the sum of the limits. This is
equivalent to the convention
$G(q)=(1/{2\pi})\int{d\omega'}\,(-i\omega'+\nu _0 k^2)^{-1}=1/2$
and gives
\begin{equation}
\Sigma_{\theta'\theta} (\omega, k)= (-1/2)
\int\frac{d{\bf q}}{(2\pi)^{d}}
\frac{D_{0} (k^{2}-({\bf k}\cdot{\bf q})^{2}/q^{2}) +
D_{0}'({\bf k}\cdot{\bf q})^{2}/q^{2}} {(q^{2}+m^{2})^{d/2+\eps/2}}.
\label{sigma3}
\end{equation}
The integration over ${\bf q}$ in (\ref{sigma3}) is performed explicitly:
\begin{mathletters}
\label{otvet}
\begin{equation}
\Sigma_{\theta'\theta} (\omega, k)= -k^{2}\,
\frac{D_{0}\,(d-1)+D_{0}'}{2d} \,  J(m),
\label{otvet1}
\end{equation}
where we have written
\begin{equation}
J(m)\equiv\int\frac{d{\bf q}}{(2\pi)^{d}}\,
\frac{1}{(q^{2}+m^{2})^{d/2+\eps/2}}
= \frac{\Gamma(\eps/2)\,m^{-\eps}}{(4\pi)^{d/2}\Gamma(d/2+\eps/2)}.
\label{otvet2}
\end{equation}
\end{mathletters}

Equations (\ref{Dyson1}), (\ref{otvet}) give an explicit exact expression
for the response function in our model; it will be used in Sec. \ref{sec:3}
for the exact calculation of the RG functions. Below we use the intermediate
expression (\ref{sigma3}). The integration of Eq. (\ref{Dyson2})  over
the frequency $\omega$ gives a closed equation for the single-time
correlator. Using (\ref{sigma3}) it can be written in the form
\begin{equation}
2\nu_0 k^{2}D(k)=C(k) +  \int\frac{d{\bf q}}{(2\pi)^{d}} \,
\frac{D_{0} (k^{2}-({\bf k}\cdot{\bf q})^{2}/q^{2}) +
D_{0}'({\bf k}\cdot{\bf q})^{2}/q^{2}} {(q^{2}+m^{2})^{d/2+\eps/2}}\,
\left[D(|{\bf k}-{\bf q}|)-D(k)\right].
\label{9}
\end{equation}
The function $C(k)$ is supposed to be analytic in $k^{2}$, which along
with the requirement that $C(k=0)=0$ (so that the Eq. (\ref{1}) has the
form of a conservation law for $\theta$) gives
\begin{equation}
C(k)=k^{2}\Psi(k)
\label{noise}
\end{equation}
with some function $\Psi(k)$, or in the coordinate representation
$C(Mr)=-\triangle\Psi(r)$, where $\Psi(r)$ vanishes rapidly for
$r\to\infty$.

In the coordinate representation, Eq. (\ref{9}) takes the form
\begin{equation}
2\nu_0 \triangle D(r)=\triangle\Psi+D_{0}(\delta_{ij}\triangle-
\partial_{i}\partial_{j}) (A_{ij}D(r))+ D_{0}'\partial_{i}\partial_{j}
 (A_{ij}D(r)),
\label{12}
\end{equation}
where we have written
\begin{equation}
A_{ij}({\bf r}) \equiv \int\frac{d{\bf q}}{(2\pi)^{d}} \,
\frac{q_{i}q_{j}\left[\exp\,({\rm i}{\bf q}\cdot{\bf r})-1\right]}
{q^{2}(q^{2}+m^{2})^{d/2+\eps/2}}.
\label{10}
\end{equation}
For $D_{0}'=0$, Eq. (\ref{12}) coincides (up to the notation) with the
well-known equation for the single-time correlator in the model
(\ref{3}) obtained in  \cite{Kraich1}.

For $0<\eps <2$,
the equations (\ref{9}), (\ref{10}) allow for the limit $m\to0$:
the possible IR divergence of the integrals at ${\bf q}=0$ is
suppressed by the vanishing of the expressions in the square brackets.
In what follows we set $m=0$. Then Eq. (\ref{10}) gives
\begin{mathletters}
\label{11}
\begin{equation}
A_{ij}({\bf r}) = -B r^{\eps}
\left[ \delta_{ij}+\eps r_{i}r_{j}/r^{2}\right],
\label{11a}
\end{equation}
\begin{equation}
B\equiv \frac{-\Gamma(-\eps/2)}
{(4\pi)^{d/2} 2^{\eps}(d+\eps) \Gamma(d/2+\eps/2)}
\label{11b}
\end{equation}
\end{mathletters}
(note that $B>0$). Using Eq. (\ref{11}) and the fact that
the function $D(r)$ depends
only on $r=|{\bf x}-{\bf x'}|$, the differential operators entering
into Eq. (\ref{12}) are represented in the form
\begin{mathletters}
\label{differ}
\begin{equation}
\partial_{i}\partial_{j} \left(A_{ij}D (r)\right) =
- B(1+\eps) r^{1-d} \partial_{r} \left(r^{(d-1)/(1+\eps)}
\partial_{r} \left(r^{\eps(d-1)/(1+\eps)} D(r) \right)\right),
\label{differ1}
\end{equation}
\begin{equation}
(\delta_{ij}\triangle- \partial_{i}\partial_{j}) (A_{ij}D (r))  =
(d-1) r^{1-d} \partial_{r} \left(r^{d-1+\eps} \partial_{r}  D (r) \right),
\label{differ2}
\end{equation}
where  $\partial_{r}\equiv \partial/\partial r$;
and for the $d$-dimensional Laplace operator one has:
\begin{equation}
\triangle \Psi(r)=r^{1-d} \partial_{r} \left(r^{d-1}
\partial_{r} \Psi(r)\right).
\label{differ3}
\end{equation}
\end{mathletters}
It then follows from (\ref{differ}) that one integration in Eq. (\ref{12})
is readily performed: one can just omit the overall ``factor''
$r^{1-d} \partial_{r} r^{d-1}$; the integration constant
is determined by the requirement that the solution have no
singularity at the origin ($r=0$):
\begin{equation}
2\nu_0 \partial_{r} D=\partial_{r}\Psi-
B(d-1)D_{0} r^{\eps}\partial_{r} D -
B (1+\eps)D_{0}' r^{-\eps(d-1)/(1+\eps)}
\partial_{r} \left(r^{\eps(d+\eps)/(1+\eps)} D \right).
\label{13}
\end{equation}
Equation (\ref{13}) is rewritten in the form
\begin{equation}
\partial_{r} \left[ \left(1+h_{0}(\Lambda r)^{\eps}
\right)^{\zeta}\,D(r) \right]=
[1+h_{0}(\Lambda r)^{\eps}]^{\zeta-1} \partial_{r} \tilde\Psi,
\label{17}
\end{equation}
where we have denoted
\begin{mathletters}
\label{15}
\begin{equation}
h_{0} \equiv B\,\frac{(d-1)+\alpha(1+\eps)}{2} ,
\end{equation}
\begin{equation}
\tilde\Psi\equiv \Psi/2\nu_{0},
\end{equation}
\end{mathletters}
and the exponent $\zeta$  has the form
\begin{mathletters}
\label{18}
\begin{equation}
\zeta = \frac{(d+\eps)\,D_{0}'}{(d-1)D_{0}+(1+\eps)\,D_{0}'},
\label{18a}
\end{equation}
so that
\begin{equation}
\zeta-1= \frac{(d-1)\,(D_{0}'-D_{0})}{(d-1)\,D_{0}+(1+\eps)\,D_{0}'}.
\label{18b}
\end{equation}
\end{mathletters}
Equation (\ref{17}) is integrated explicitly; the integration constant is
found from the requirement that the solution vanish at infinity
(including the special case $h_{0}=0$):
\begin{equation}
D(r)= \frac{-1}{[1+h_{0}(\Lambda r)^{\eps}]^{\zeta}}
\int^{\infty}_{r} dy\, [1+h_{0}(\Lambda y)^{\eps}]^{\zeta-1}
\partial_{y} \tilde\Psi(y).
\label{20}
\end{equation}
For $D_{0}'=0$ (so that $\alpha=\zeta =0$), the
expression (\ref{20}) reduces (up to the notation)
to the well known solution for the purely solenoidal velocity field
obtained in \cite{Kraich1}. Dimensionality considerations give
$\Psi(r)=M^{-2}\psi(Mr)$ with some dimensionless function $\psi$,
see (\ref{noise}), so that (\ref{20}) can be rewritten as
\begin{equation}
D(r)=\frac{-1}{2\nu_0 M^{2}[1+h_{0}(\Lambda r)^{\eps}]^{\zeta}}
\int^{\infty}_{Mr} dy [1+h_{0}(\Lambda y/M)^{\eps}]^{\zeta-1}
\partial_{y} \psi(y).
\label{21}
\end{equation}

We are interested in the asymptotic form of the correlator $D(r)$ and
the structure function $S_{2}\propto D(0)-D(r)$ in the convective range
$\Lambda >>1/r>>M$,  where $\Lambda$  is determined by (\ref{Lambda}).
From (\ref{21}) it then follows
\begin{mathletters}
\label{22}
\begin{equation}
D(r=0)\simeq C\, \nu_0^{-1} \,
M^{-2-\epsilon(\zeta-1)}\, \Lambda ^{\epsilon(\zeta-1)},
\label{22a}
\end{equation}
where we have used the definitions (\ref{Lambda}) and (\ref{15}),
and $C$ is completely dimensionless factor independent of $r$, $M$
and $\Lambda$:
\begin{equation}
C\equiv \frac{-h_{0}^{\zeta-1}}{2}  \int_{0}^{\infty} dy
y^{\eps(\zeta-1)} \partial_{y}\psi(y).
\label{22b}
\end{equation}
For the correlator $D(r)$ in the region $\Lambda >>1/r>>M$ one obtains:
\begin{equation}
D(r)\simeq h_{0}^{-\zeta}  (\Lambda r)^{-\eps\zeta} D(r=0).
\label{23}
\end{equation}
\end{mathletters}
It follows from (\ref{22}) that $D(r)$ differs from $D(0)$ by the
factor $\propto(\Lambda r)^{-\eps\zeta}<<1$. Therefore, the leading
contribution to the structure function $S_{2}\propto D(0)-D(r)$ in the
convective range is given by the constant term $D(0)$, while the $r$
dependent contribution determines only a vanishing correction. Then
the comparison of the expression (\ref{as2}) for $n=1$ with the exact
result (\ref{22a}) gives $\Delta_{2}=-2+\eps-\eps\zeta$, which along
with Eq. (\ref{18a}) leads to the exact expression (\ref{exp2}) for
the critical dimension $\Delta_{2}$, announced in the Introduction.

The expressions (\ref{22}) simplify for $d=1$
(and for $D_{0}=D_{0}'$ and any $d$), when $\zeta=1$, see (\ref{18b}):
\begin{mathletters}
\label{24}
\begin{equation}
D(r)\propto \nu_0^{-1} M^{-2} (\Lambda r)^{-\eps},
\label{24a}
\end{equation}
\begin{equation}
D(r=0)\propto \nu_0^{-1} M^{-2} .
\label{24b}
\end{equation}
\end{mathletters}

The expression (\ref{24a}) agrees\footnote{
In the previous version of this paper, it was erroneously stated that
the expression (\ref{24a}) disagrees with the result \cite{VM}.
The authors are thankful to M.~Vergassola for pointing this out
to them.} with the result obtained
in \cite{VM} directly for $d=1$. In the language of the papers
\cite{Falk1,Falk2,GK,BGK,VM}, the leading non-universal term in
(\ref{24a}) is related to a nontrivial zero mode of the differential
operator entering into Eq. (\ref{12}). We note that the anomalous
scaling for the pair correlator with the exactly known exponent was
established previously in \cite{MHD} on the example of a passively
advected magnetic field, as a result of the existence of a nontrivial
zero mode of the corresponding differential operator. We also note
that for the purely solenoidal case, the analogous zero mode is
independent of $r$ and cancels out in the structure function, so
that the IR behavior of the latter is determined by the universal
correction term $r^{2-\eps}$.

In the subsequent Sections, the asymptotic expressions
(\ref{22}) will be generalized to the case of higher-order
correlators and structure functions.

\section{Renormalization, RG functions, and RG equations}
\label {sec:3}

The analysis of the UV divergences in a field theoretical model
is based on the analysis of canonical dimensions. Dynamical models
of the type (\ref{action}), in contrast to static models, are
two-scale, i.e., to each quantity $F$ (a field or a parameter
in the action functional) one can assign two independent canonical
dimensions, the momentum dimension $d_{F}^{k}$ and the frequency
dimension $d_{F}^{\omega}$, determined from the natural
normalization conditions $d_k^k=-d_{\bf x}^k=1$, $d_k^{\omega }
=d_{\bf x}^{\omega }=0$, $d_{\omega }^k=d_t^k=0$,
$d_{\omega }^{\omega }=-d_t^{\omega }=1$, and from the requirement
that each term of the action functional be dimensionless (with
respect to the momentum and frequency dimensions separately),
see e.g. \cite{UFN,Pismak,JETP}. Then, based on $d_{F}^{k}$ and
$d_{F}^{\omega}$,
one can introduce the total canonical dimension
$d_{F}=d_{F}^{k}+2d_{F}^{\omega}$ (in the free theory,
$\partial_{t}\propto\triangle$).

The dimensions for the model (\ref{action}) are given in
Table \ref{table1}, including renormalized parameters,
which will be considered later on.
From Table \ref{table1} it follows that the model is
logarithmic (the coupling constant $g_{0}$ is dimensionless)
at $\eps=0$, and the UV divergences have the form of
the poles in $\eps$ in the Green functions.
The total dimension $d_{F}$ plays in the theory of
renormalization of dynamical models the same role as does
the conventional (momentum) dimension in static problems.
The canonical dimensions of an arbitrary
1-irreducible Green function $\Gamma = \langle\Phi \dots \Phi
\rangle _{\rm 1-ir}$ are given by the relations
\begin{mathletters}
\label{delta}
\begin{equation}
d_{\Gamma }^k=d- N_{\Phi}d_{\Phi},
\label{deltaa}
\end{equation}
\begin{equation}
d_{\Gamma }^{\omega }=1-N_{\Phi }d_{\Phi }^{\omega },
\label{deltab}
\end{equation}
\begin{equation}
d_{\Gamma }=d_{\Gamma }^k+2d_{\Gamma }^{\omega }=
d+2-N_{\Phi }d_{\Phi},
\label{deltac}
\end{equation}
\end{mathletters}
where $N_{\Phi}=\{N_{\theta},N_{\theta'},N_{\bf v}\}$ are the
numbers of corresponding fields entering into the function
$\Gamma$, and the summation over all types of the fields is
implied.
The total dimension $d_{\Gamma}$ is the formal index of the
UV divergence. Superficial UV divergences, whose removal requires
counterterms, can be present only in those functions $\Gamma$ for
which $d_{\Gamma}$ is a non-negative integer.

Analysis of the divergences should be based on the following auxiliary
considerations, see \cite{RG,UFN,Pismak,JETP}:

(i) From the explicit form of the vertex and bare propagators in
the model (\ref{action}) it follows that $N_{\theta'}- N_{\theta}=2N_{0}$
for any 1-irreducible Green function, where $N_{0}\ge0$
is the total number of the bare propagators $\langle \theta \theta
\rangle _0$ entering into the function (obviously, no diagrams
with $N_{0}<0$ can be constructed). Therefore, the difference
$N_{\theta'}- N_{\theta}$ is an even non-negative integer for
any nonvanishing function.

(ii) If for some reason a  number of external momenta occur as an
overall factor in all the diagrams of a given Green function, the
real index of divergence $d_{\Gamma}'$ is smaller than $d_{\Gamma}$
by the corresponding number of unities (the Green function requires
counterterms only if $d_{\Gamma}'$  is a non-negative integer).

In the model (\ref{action}), the derivative
$\partial$ at the vertex $\theta'\partt({\bf v}\theta)$
can be moved onto the field $\theta'$ using the integration
by parts, which decreases the real index of divergence:
$d_{\Gamma}' = d_{\Gamma}- N_{\theta'}$.
The field $\theta'$ enters into the counterterms only
in the form of the derivative $\partial\theta'$.

(iii) A great deal of diagrams in the model (\ref{action})
contain effectively closed circuits of retarded
propagators $\langle\theta\theta'\rangle_{0}$  and therefore
vanish. For example, all the nontrivial diagrams of the
1-irreducible function $\langle\theta\theta'{\bf v}\rangle_{1-ir}$
vanish.

From the dimensions in Table \ref{table1} we find
$d_{\Gamma} = d+2 - N_{\bf v} + N_{\theta}- (d+1)N_{\theta'}$
and $d_{\Gamma}'=(d+2)\,(1-N_{\theta'})-N_{\bf v}+ N_{\theta}$.
From these expressions it follows that for any $d$,
superficial divergences can only exist in the 1-irreducible functions
with $N_{\theta'}=1$,  $N_{\bf v}=N_{\theta}=0 $
($d_{\Gamma}=1$, $d_{\Gamma}'=0$),
$N_{\theta'}=N_{\bf v}=N_{\theta}=1$
($d_{\Gamma}=1$, $d_{\Gamma}'=0$),  and
$N_{\theta'}=N_{\theta}=1$, $N_{\bf v}=0$
($d_{\Gamma}=2$, $d_{\Gamma}'=1$)
(we recall that $N_{\theta}\le N_{\theta'}$, see (i) above).
However, no diagrams can be constructed for the first of these
functions, while for the second function,
all the nontrivial diagrams vanish (see (iii) above).
As in the case of the purely solenoidal field \cite{RG},
we are left with the only superficially
divergent function $\langle\theta'\theta\rangle_{1-ir}$;
the corresponding counterterm necessarily contains the factor of
$\partial\theta'$  and is therefore reduced to $\theta'\triangle\theta$.
Introduction of this counterterm is reproduced by the multiplicative
renormalization of the parameters $g_{0},\nu_0$ in the action
functional (\ref{action}) with the only
independent renormalization constant $Z_{\nu}$:
\begin{mathletters}
\label{reno}
\begin{equation}
\nu_0=\nu Z_{\nu},
\label{reno1}
\end{equation}
\begin{equation}
g_{0}=g\mu^{\eps}Z_{g},
\label{reno2}
\end{equation}
\begin{equation}
Z_{g}=Z_{\nu}^{-1}.
\label{reno3}
\end{equation}
\end{mathletters}
Here $\mu$ is the renormalization mass in the minimal subtraction
scheme (MS), which we always use in what follows, $g$ and $\nu$
are renormalized analogues of the bare parameters $g_{0}$ and $\nu_0$,
and $Z=Z(g,\alpha,\eps,d)$ are the renormalization constants. Their
relation in (\ref{reno3}) results from the absence of renormalization
of the contribution with $D_{v}$ in (\ref{action}), so that
$D_{0}\equiv g_{0}\nu_0 = g\mu^{\eps} \nu$. No renormalization
of the fields and the parameters $m,M,\alpha$ is required,
i.e., $Z_{\Phi}=1$ for all $\Phi$ and  $m_{0}=m$, $Z_{m}=1$, etc.
The renormalized action functional has the form
\begin{equation}
S_{ren}(\Phi)=\theta' D_{\theta}\theta' /2+
\theta' \left[ - \partial_{t}\theta -\partt({\bf v} \theta)
+ \nu Z_{\nu}\triangle \theta \right]
-{\bf v} D_{v}^{-1} {\bf v}/2,
\label{renact}
\end{equation}
where the contribution with $D_{v}$ is expressed in renormalized
parameters using (\ref{reno}).

The relation $ S(\Phi,e_{0})=S_{ren}(\Phi,e,\mu)$ (where $e_{0}$
is the complete set of bare parameters, and $e$ is the set of renormalized
parameters) for the generating functional $W(A)$ in (\ref{field})
yields $ W(A,e_{0})=W_{ren}(A,e,\mu)$. We use $\widetilde{\cal D}_{\mu}$
to denote the differential operation $\mu\partial_{\mu}$ for fixed
$e_{0}$ and operate on both sides of this equation with it. This
gives the basic RG differential equation:
\begin{mathletters}
\label{RG}
\begin{equation}
{\cal D}_{RG}\,W_{ren}(A,e,\mu)  = 0,
\label{RG1}
\end{equation}
where ${\cal D}_{RG}$ is the operation $\widetilde{\cal D}_{\mu}$
expressed in the renormalized variables:
\begin{equation}
{\cal D}_{RG}\equiv {\cal D}_{\mu} + \beta(g)\partial_{g}
-\gamma_{\nu}(g){\cal D}_{\nu},
\label{RG2}
\end{equation}
\end{mathletters}
where we have written ${\cal D}_{x}\equiv x\partial_{x}$ for
any variable $x$, and the RG functions (the $\beta$ function and
the anomalous dimension $\gamma$) are defined as
\begin{mathletters}
\label{RGF}
\begin{equation}
\gamma_{\nu}(g)\equiv\widetilde{\cal D}_\mu \ln Z_{\nu},
\label{RGF1}
\end{equation}
\begin{equation}
\beta(g)\equiv \widetilde{\cal D}_\mu g= g\,(-\eps + \gamma_{\nu}).
\label{RGF2}
\end{equation}
\end{mathletters}
The relation between $\beta$ and $\gamma$ in (\ref{RGF2}) results from the
definitions and the relation (\ref{reno3}).

The renormalization constant $Z_{\nu}$ is found from the requirement
that the 1-irreducible function $\langle\theta'\theta\rangle_{1-ir}$
expressed in renormalized variables be UV finite (i.e.,
be finite for $\eps\to0$). This requirement determines $Z_{\nu}$ up to
an UV finite contribution; the latter is fixed by the choice of a
renormalization scheme. In the MS scheme all renormalization constants
have the form ``1 + only poles in  $\eps$.''  The function
$G^{-1}=\langle\theta'\theta\rangle_{1-ir}$ in our model is known exactly,
see Eqs. (\ref{Dyson1}), (\ref{otvet}).  Let us substitute (\ref{reno})
into Eqs. (\ref{Dyson1}), (\ref{otvet})
and choose $Z_{\nu}$ to cancel the pole in $\eps$
in the integral $J(m)$. This gives:
\begin{equation}
Z_{\nu}= 1 - g\, C_{d} \frac{d-1+\alpha}{2d\eps},
\label{Z}
\end{equation}
where we have written
$C_{d} \equiv  {S_{d}}/{(2\pi)^{d}}$
and $S_d\equiv 2\pi ^{d/2}/\Gamma (d/2)$
is the surface area of the unit sphere in $d$-dimensional space.
Note that the result (\ref{Z}) is exact, i.e., it has no corrections
of order $g^{2}$, $g^{3}$, and so on; this is a consequence of the fact
that the single-loop approximation (\ref{otvet})
for the response function is exact. Note also that for $\alpha=0$
(\ref{Z}) coincides with the exact expression for $Z_{\nu}$ in the
``incompressible'' case obtained in \cite{RG}.

For the anomalous dimension $\gamma_{\nu}(g)\equiv
\widetilde{\cal D}_\mu \ln Z_{\nu}
=\beta(g)\partial_{g}\ln Z_{\nu}$  from the relations (\ref{RGF2})
and (\ref{Z}) one obtains:
\begin{equation}
\gamma_{\nu}(g)=\frac{-\eps {\cal D}_g \ln Z_{\nu}}{1-{\cal D}_g \ln Z_{\nu}}=
g\, C_{d} \frac{d-1+\alpha}{2d}.
\label{gammanu}
\end{equation}
From (\ref{RGF2}) it then follows that the RG equations of the model
have an IR stable fixed point [$\beta(g_{*})=0$, $\beta'(g_{*})>0$]
with the coordinate
\begin{equation}
g_{*}= \frac{2d\eps}{C_{d}(d-1+\alpha)}.
\label{fixed}
\end{equation}
The fixed point is degenerated: its coordinate $g_{*}$ depends
continuously on the parameter $\alpha=D_{0}'/D_{0}$.\footnote{ Formally,
$\alpha$ can be treated as the second coupling constant. The corresponding
beta-function $\beta_{\alpha}\equiv\widetilde{\cal D}_\mu{\alpha}$ vanishes
identically owing to the fact that $\alpha$ is not renormalized. Therefore,
the equation $\beta_{\alpha}=0$ gives no additional constraint on the values
of the parameters $g,\alpha$ at the fixed point.} The value
of $\gamma_{\nu}(g)$ at the fixed point is also found  exactly:
\begin{equation}
\gamma_{\nu}^{*} \equiv \gamma_{\nu}(g_*)= \eps.
\label{27}
\end{equation}

The solution of the RG equations on the example of the stochastic
hydrodynamics is discussed in detail in Refs.\cite{UFN,JETP};
see also \cite{RG} for the case of the model (\ref{3}); below we confine
ourselves to the only information we need.

In general, if some quantity $F$ (a parameter, a field or composite
operator) is renormalized multiplicatively, $F=Z_{F}F_{ren}$ with
certain renormalization constant $Z_{F}$, its critical dimension
is given by the expression (cf.\cite{UFN,Pismak,JETP}):
\begin{equation}
\Delta[F]\equiv\Delta_{F} = d_{F}^{k}+ \Delta_{\omega}
d_{F}^{\omega}+\gamma_{F}^{*},
\label{32B}
\end{equation}
where $d_{F}^{k}$ and $d_{F}^{\omega}$ are the corresponding canonical
dimensions, $\gamma_{F}^{*}$ is the value of the anomalous dimension
$\gamma_{F}(g)\equiv \widetilde{\cal D}_\mu \ln Z_{F}$  at the fixed
point, and $\Delta_{\omega}=-2+\gamma^{*}_{\nu}=-2+\eps$ is the critical
dimension of frequency. The critical dimensions of the fields $\Phi$ in
our model are found exactly; they are independent of the parameter $\alpha$
and coincide with their analogues in the model (\ref{3}), cf.\cite{RG}:
\begin{mathletters}
\label{33}
\begin{equation}
\Delta_{\bf v}=1-\eps,
\label{33a}
\end{equation}
\begin{equation}
\Delta_{\theta} = -1+\eps/2,
\label{33b}
\end{equation}
\begin{equation}
\Delta_{\theta'} = d+1-\eps/2
\label{33c}
\end{equation}
\end{mathletters}
(we recall that the fields in the model (\ref{action}) are
not renormalized and therefore $\gamma_{\Phi}=0$ for all $\Phi$).

Let $G(r)=\langle F_{1}(x)F_{2}(x')\rangle$ be a single-time
two-point quantity, for example, the pair correlation function
of the primary fields $\Phi\equiv\{\theta, \theta',{\bf v}\}$
or some multiplicatively renormalizable composite operators.
The existence of the IR stable fixed point implies that in the
IR asymptotic region $\Lambda r>>1$ and any fixed $Mr$ the function
$G(r)$ is found in the form
\begin{equation}
G(r) \simeq  \nu_{0}^{d_{G}^{\omega}}\, \Lambda^{d_{G}}
(\Lambda r)^{-\Delta_{G}}\, \xi(Mr),
\label{55}
\end{equation}
with certain, as yet unknown, scaling function $\xi$ of the critically
dimensionless argument $Mr$.
The canonical dimensions $d_{G}^{\omega}$, $d_{G}$ and the
critical dimension $\Delta_{G}$ ot the function $G(r)$ are equal
to the sums of the corresponding dimensions of the quantities
$F_{i}$.

Now let us turn to the composite operators of the form $\theta^{n}(x)$
entering into the structure functions (\ref{struc})
and the correlators  (\ref{as1}).

In general, counterterms to a given operator $F$ are
determined by all possible 1-irreducible Green functions
with one operator $F$ and arbitrary number of primary fields,
$\Gamma=\langle F(x) \Phi(x_{1})\dots\Phi(x_{2})\rangle_{1-ir}$.
The total canonical dimension (formal index of divergence)
for such functions is given by
\begin{equation}
d_\Gamma = d_{F} - N_{\Phi}d_{\Phi},
\label{index}
\end{equation}
with the summation over all types of fields entering into
the function. For superficially divergent diagrams, the real index
$d_\Gamma'=d_\Gamma-N_{\theta'}$ is a non-negative integer.
From Table I and Eq. (\ref{index}) for the operators $\theta^{n}(x)$
we obtain $d_{F}=-n$,
$d_\Gamma = -n+N_{\theta}-N_{\bf v} -(d+1)N_{\theta'}$,
and $d_\Gamma' = -n+N_{\theta}-N_{\bf v} -(d+2)N_{\theta'}$.
From the analysis of the diagrams it follows that the total
number of the fields $\theta$ entering into the function
$\Gamma$ can never exceed the number of the fields $\theta$
in the operator $\theta^{n}$ itself, i.e., $N_{\theta}\le n$.
Therefore, the divergence can only exist in the functions
with $N_{\bf v}=0$, $N_{\theta'}=0$, and arbitrary value of
$n=N_{\theta}$, for which $d_\Gamma=d_\Gamma'=0$ and the
corresponding counterterm has the form $\theta^{n}$. It then
follows that the operator $\theta^{n}$ is renormalized multiplicatively,
$\theta^{n}=Z_{n}[\theta^{n}]_{ren}$.

Note an important difference between the case of a purely transversal
velocity field (\ref{3}) and the general case (\ref{4}).
In the first case, the derivative $\partt$ at the vertex
can be moved onto the field $\theta$ owing to the transversality
of the velocity field,
$\theta'\partt({\bf v}\theta)=\theta'({\bf v}\partt)\theta$.
This reduces the real index $d_\Gamma'$ by at least one unity,
so that $d_\Gamma'$ becomes strictly negative, see \cite{RG}.
This means that the operator $\theta^{n}$ requires no
counterterms at all, i.e., it  is in fact UV
finite, $Z_{n}=1$.\footnote{ This ``non-renormalization''  result
can be interpreted as the fact that the scalar field remains a
continuous function even in the limit $\nu_0\to0$  or equivalently
$\Lambda\to\infty$. The nontrivial UV renormalization of the
monomials $(\partial\theta\partial\theta)^{n}$  \cite{RG} points to
the fact that the scalar field is not differentiable, i.e., its
gradients exist only as distributions. One of the authors (N.V.A)
is thankful to G.L.Eyink for pointing this out to him, see also
Refs.\cite{Eyink,Xin,Xin2}.} It then follows that the critical dimension
of $\theta^{n}(x)$ in the model (\ref{3})
is simply given by the expression  (\ref{32B})
with no correction from $\gamma_{F}^{*}$ and is therefore reduced
to the sum of the critical dimensions of the factors \cite{RG}:
\begin{equation}
\Delta_{n}\equiv\Delta [\theta^{n}] = n\Delta[\theta] =n (-1+\eps/2).
\label{simple}
\end{equation}
In the general case (\ref{4}), the constants $Z_{n}$ are nontrivial,
and the simple relation (\ref{simple}) is no longer valid.
The two-loop calculation of the constants $Z_{n}$ is explained in detail
in Sec. \ref{sec:5}, and here we only give the two-loop result
for the critical dimensions $\Delta_{n}$ in the model (\ref{action}):
\begin{eqnarray}
\Delta_{n}= n(-1+\eps/2)-\frac{\alpha n(n-1)d\eps}{2(d-1+\alpha)}+
\frac{\alpha(\alpha-1) n(n-1)(d-1)\eps^{2}}{2(d-1+\alpha)^{2}} +
\nonumber    \\
\nonumber    \\
+\frac{\alpha^{2} n(n-1)(n-2)d\,h(d)\eps^{2}}{4(d-1+\alpha)^{2}}+
O(\eps^{3}),
\label{Dn}
\end{eqnarray}
where we have denoted
\begin{equation}
h(d)\equiv \sum_{k=0}^{\infty} \frac{k!}{4^{k}(d/2+1)\dots(d/2+k)}=
F(1,1;d/2+1;1/4),
\label{hyper}
\end{equation}
and $F(\dots)$ is the hypergeometric series, see \cite{Gamma}.

In the special case $n=2$ one obtains from (\ref{Dn}):
\begin{equation}
\Delta_{2}= -2-\frac{\eps(d-1)(\alpha-1)}{(d-1+\alpha)}+
\frac{\eps^{2}(d-1)\alpha(\alpha-1)}{(d-1+\alpha)^{2}}+O(\eps^{3}).
\label{D2}
\end{equation}

Expression (\ref{Dn}) is simplified for any integer value
of $d$ owing to the fact that the series in (\ref{hyper})
reduces then to a finite sum, see \cite{Gamma}:
\begin{mathletters}
\label{Gamma1}
\begin{equation}
h(d)=2d\left[(-3)^{d/2-1}\ln(4/3) +
\sum_{k=2}^{d/2} \frac{(-3)^{k-2}} {d/2-k+1}\right]
\label{Gamma4}
\end{equation}
for any even value of $d$ and
\begin{equation}
h(d)=2d\left[(-1)^{(d-1)/2}\cdot 3^{d/2-2}\cdot\pi+2\,
\sum_{k=1}^{(d-1)/2} \frac{(-3)^{(d-1)/2-k}} {2k-1}\right]
\label{Gamma3}
\end{equation}
\end{mathletters}
for any odd value of $d$, which gives $h(d)=2\pi/(3\sqrt 3)$ for $d=1$,
$h(d)=4\ln(4/3)$ for $d=2$, and $h(d)=12-2\pi\sqrt 3$ for $d=3$.
(We note that for $d=1$ and
$d=2$ the sums in (\ref{Gamma1}) contain no terms).
The case of a purely potential velocity field is obtained for
$D_{0}'=\const$, $D_{0}=0$ or, equivalently, $\alpha\to\infty$,
$g_{0}'\equiv g_{0}\alpha =\const$. From (\ref{fixed}) it then follows
that at the fixed point $g'_{*}=2d\eps/C_{d} $;
the values of the critical dimensions $\Delta_{n}$
are obtained simply by taking the limit $\alpha\to\infty$
in the expressions (\ref{Dn}),  (\ref{D2}) and have the form
\begin{eqnarray}
\Delta_{n}= n(-1+\eps/2) - n(n-1)d\eps/2 + n(n-1)(d-1)\eps^{2}/2 +
\nonumber \\
+n(n-1)(n-2)\,  h(d)\,d \eps^{2}/4 +O(\eps^{3}) .
\label{dip}
\end{eqnarray}
In the special case $d=1$ one obtains
\begin{equation}
\Delta_{n}=-n+n\eps -n^{2}\eps/2 +
n(n-1)(n-2)\eps^{2}\pi/(6\sqrt 3)+O(\eps^{3}).
\label{d1}
\end{equation}

For the pair correlators of the operators  $\theta^{n}$
we obtain from Table I and Eqs. (\ref{55}), (\ref{Dn}):
\begin{equation}
\langle\theta^{n}(x)\theta^{p}(x')\rangle=\nu_0^{-(n+p)/2}\,
\Lambda^{-(n+p)}
(\Lambda r)^{-\Delta_{n}-\Delta_{p}} \xi_{n,p}(Mr),
\label{as11}
\end{equation}
with the dimensions $\Delta_{n}$ given in (\ref{Dn}) and certain
scaling functions $\xi_{n,p}(Mr)$ (for odd $n+p$ they vanish).
We recall that the representation (\ref{as11}) holds for
$\Lambda r>>1$ and any fixed $Mr$; the behavior of the functions
$\xi_{n,p}(Mr) $ for $Mr<<1$ (convective range) is studied in the
subsequent section.

\section{Operator product expansion and anomalous scaling}
\label {sec:4}

The representation (\ref{as11}) for any functions $\xi_{n,p}(Mr)$
correspond to IR scaling in the region $\Lambda r>>1$ and any fixed $Mr$
with definite critical dimensions $\Delta_{n}$ given in (\ref{Dn}).
The expressions (\ref{as1}) should be understood as certain additional
statements about the explicit form of the asymptotic behavior
of the functions $\xi_{n,p}(Mr)$ for $Mr\to0$.
The form of the scaling functions $\xi_{n,p}(Mr)$ in the representation
(\ref{as11}) is not determined by the RG equations themselves;
these functions can be calculated in the form of series in $\eps$.
However, this $\eps$ expansion is not suitable for the analysis of their
behavior for $Mr\to0$, because the actual expansion parameter appears
to be $\eps\ln(Mr)$  rather than $\eps$ itself, cf.\cite{RG,UFN,JETP}.
In contrast to the ``large UV logarithms'' $\ln(\Lambda r)$, the
summation of these ``large IR logarithms''  is not performed
automatically by the solution of the RG equations.

In the theory of critical phenomena, the asymptotic form of
scaling functions for  $M\to0$ is studied using the well
known Wilson operator product expansion (OPE), see e.g.
 \cite{Zinn}; the analogue of $L\equiv M^{-1}$
is there the correlation length $r_{c}$.
This technique is also applied to the theory of
turbulence, see e.g. \cite{RG,UFN,JETP}.

According to the OPE, the single-time product $F_{1}(x)F_{2}(x')$
of two renormalized operators at
${\bf x}\equiv ({\bf x} + {\bf x'} )/2 = {\const}$, and
${\bf r}\equiv {\bf x} - {\bf x'}\to 0$
has the representation
\begin{equation}
F_{1}(x)F_{2}(x')=\sum_{\alpha}C_{\alpha} ({\bf r})
F_{\alpha}({\bf x,t}) ,
\label{OPE}
\end{equation}
in which the functions $C_{\alpha}$  are the Wilson coefficients
regular in $M^{2}$ and  $ F_{\alpha}$ are all possible
renormalized local composite operators allowed by symmetry, with
definite critical dimensions $\Delta_{\alpha}$.

The renormalized correlator $\langle F_{1}(x)F_{2}(x') \rangle$
is obtained by averaging (\ref{OPE}) with the weight
$\exp S_{R}$, the quantities  $\langle F_{\alpha}\rangle$
appear on the right hand side. Their asymptotic behavior
for $M\to0$ is found from the corresponding RG equations and
has the form
\begin{equation}
\langle F_{\alpha}\rangle \propto  M^{\Delta_{\alpha}}.
\label{2.45}
\end{equation}
From the operator product expansion (\ref{OPE}) we therefore
find the following expression  for the scaling function
$\xi(Mr)$ in the representation (\ref{55}) for the correlator
$\langle F_{1}(x)F_{2}(x') \rangle$:
\begin{equation}
\xi(Mr)=\sum_{\alpha}A_{\alpha}\, (Mr)^{\Delta_{\alpha}},
\label{2.46}
\end{equation}
with coefficients $A_{\alpha}=A_{\alpha}(Mr)$, which are regular
in  $(Mr)^{2}$,
generated by the Wilson coefficients $C_{\alpha}$ in (\ref{OPE}).

We note that for a Galilean invariant product $F_{1}(x)F_{2}
(x')$, the right hand side of Eq. (\ref{OPE})
can involve any Galilean invariant operator, including tensor
operators, whose indices are contracted with the analogous
indices of the coefficients $C_{\alpha}$. Without loss of
generality, it can be assumed that the expansion is made in
irreducible tensors,
so that only scalars contribute to the correlator
$\langle F_{1}F_{2}\rangle$  because the averages
$\langle F_{\alpha}\rangle$  for non-scalar irreducible
tensors vanish. For the same reason, the contributions to the
correlator from all operators of the form $\partial F$  with
external derivatives vanish owing to translational invariance.

The leading contributions for  $Mr\to0$
are those with the smallest dimension $\Delta_{\alpha}$,
and in the $\eps$ expansions they are those with the smallest
$d_{\alpha}\equiv d[F_{\alpha}]$  for $\eps=0$.
In the standard model $\phi^{4}$ of the theory of critical
behavior one has $\Delta_{\alpha} = n_{\alpha} + O(\eps)$,
where $n_{\alpha}\ge0$  is the total number of fields and
derivatives in $F_{\alpha}$.
The operator $F=1$  has the smallest value $n_{\alpha}=0$,
and it gives a contribution to (\ref{2.46})  which is regular
in  $(Mr)^{2}$ and has a finite limit as $Mr\to0$.
The first nontrivial contribution is generated by the operator
$\phi^{2}$ with $n_{\alpha}=2$. It has the form  $(Mr)^{2+O(\eps)}$
and only determines a correction, vanishing at $Mr\to0$,
to the leading term generated by the operator $F=1$.

The distinguishing feature of the models describing turbulence
is the existence of ``dangerous'' composite operators with {\it negative}
critical dimensions \cite{RG,UFN,JETP}.  The contributions of the
dangerous operators into the operator product expansions lead to
a singular behavior of the scaling functions on $M$ for $Mr\to0$.
It is obvious from (\ref{Dn}) that all the operators $\theta^{n}$
in the model (\ref{action}) are dangerous at least for small $\eps$,
and the spectrum of their critical dimensions is unbounded from below.
If all these operators contributed to the OPE like (\ref{OPE}),
the analysis of the small $M$ behavior would imply the summation
of their contributions. Such a summation is indeed required for the
case of the different-time correlators in the stochastic Navier--Stokes
equation, and it establishes the substantial dependence of the
correlators on $M$ and their superexponential decay as the time
differences increase, see \cite{UFN,JETP}. Fortunately, the problem
simplifies for the model (\ref{action}).

From the analysis of the diagrams it follows
that the number of the fields $\theta$ in the operator $F_{\alpha}$
entering into the right hand sides of the expansions (\ref{OPE})
can never exceed the total number of the  fields $\theta$
in their left hand sides. Therefore, only finite number of
operators $\theta^{n}$ contribute to each operator product
expansion, and the asymptotic form  of the scaling functions
is simply determined by the operator $\theta^{n}$ with the lowest
critical dimension, i.e., with the largest possible number
of the fields $\theta$.  For the scaling functions  $\xi_{n,p}(Mr)$
entering into the expressions (\ref{as11}) this gives
$\xi_{n,p}(Mr)\propto (Mr)^{\Delta_{n+p}}$,
which lead to the asymptotic expression (\ref{as1})
announced in the Introduction.

It is noteworthy that the set of the operators  $\theta^{n}$
is ``closed with respect to the fusion'' in the sense that
the leading term in the OPE for the pair correlator
$ \langle \theta^{n}\theta^{p} \rangle$ is given by the operator
$ \theta^{n+p}$ from the same family with the summed
index $n+m$. This fact along with the inequality
\begin{equation}
\Delta_{n}+\Delta_{p} >  \Delta_{n+p},
\label{frac}
\end{equation}
which is obvious from the explicit expression (\ref{Dn}) for
small values of $\eps$, can be interpreted as the statement
that the correlations of the scalar field $\theta$ in the model
(\ref{action}) exhibit multifractal behavior,
see \cite{DL,Eyink1,Hol}. In the case of the solenoidal
velocity field, the dimension $\Delta_{n}$ becomes linear in $n$,
see (\ref{simple}), and the relation (\ref{as1}) reduces to
the so-called ``gap scaling'' (see \cite{DL}). In this case,
the nontrivial multifractal behavior manifests itself in the
correlations of the dissipation rate rather than in the
correlations of the field itself, see \cite{RG}.

Now let us turn to the structure functions (\ref{struc}) in
the convective range $\Lambda r>>1$, $Mr<<1$. From the expression
(\ref{as1}) it follows
\begin{equation}
S_{2n} \simeq \nu_0^{-n}\, \Lambda^{-2n}\,(M/\Lambda)^{\Delta_{2n}}
\left[1+
\sum_{\scriptstyle k+p=2n \atop \scriptstyle k,p\ne0}
c_{kp} (\Lambda r)^{\Delta_{2n}-\Delta_{k}-\Delta_{p}}
\right],
\label{struc2}
\end{equation}
where the coefficients $c_{kp}$ are independent of the scales $\Lambda$,
$M$ and the separation $r$. It is obvious from the inequality
(\ref{frac}) that  all the contributions in the sum
in (\ref{struc2}) vanish in the region
$\Lambda r>>1$, so that the leading terms of the structure
functions do not depend on $r$ and are given by the Eq. (\ref{as2}).

The comparison of the expressions (\ref{as1}) and (\ref{as2})
for $k=p=1$ with the exact results (\ref{23}) and (\ref{22a})
gives $\Delta_{2}=-2+\eps-\eps\zeta$, which along with Eq. (\ref{18a})
leads to the exact expression (\ref{exp2}) for the critical dimension
$\Delta_{2}$, announced in the Introduction.
We note that the expression (\ref{D2}) for $\Delta_{2}$
obtained within the RG approach is in agreement with the
corresponding terms of the expansion in $\eps$ of the exact
exponent (\ref{exp2}) for all  $d$ and $\alpha$.
We also note that (\ref{23}) is consistent with the
exact RG result $\Delta_{\theta}=-1+\eps/2$, see (\ref{33b}).

It is seen from (\ref{struc2}) that the IR behavior for the
structure functions is determined by the contributions
of the composite operators $\theta^{n}$ to the corresponding OPE.
The operators $\theta^{n}$ obviously do not appear in the na\"{\i}ve
Taylor expansions of the structure functions (\ref{struc}) for
$r\to0$: the Taylor expansion for the function $S_{2n}$ starts
with the monomial $(\partial_{i}\theta\partial_{i}\theta)^{n}$.
However, the operators entering into operator product expansions
are not only those which appear in the Taylor expansions, but also
all possible operators which admix to them in renormalization.
One can easily check that all the monomials $\theta^{2p}$  with $p\le n$
admix to $(\partial_{i}\theta\partial_{i}\theta)^{n}$ in renormalization.
As a result, their contributions appear in the OPE for the structure
functions and dominate their IR asymptotic behavior.

The situation changes if the velocity field is purely solenoidal,
with the correlator given in (\ref{3}). In this case, the field
$\theta$ enters into the vertex in the form of a derivative,
$\theta'\partt({\bf v}\theta)=\theta'({\bf v}\partt)\theta$,
and therefore only derivatives of $\theta$ can appear in the
counterterms to the monomials
$(\partial_{i}\theta\partial_{i}\theta)^{n}$.
Hence, the operators of the form $\theta^{n}$ cannot admix in
renormalization to the monomials
$(\partial_{i}\theta\partial_{i}\theta)^{n}$
and cannot appear in the OPE for the structure functions (\ref{struc}).
This means that the contributions of the operators $\theta^{n}$ to the
pair correlators (\ref{as1}) cancel out in the structure functions,
and the IR behavior of the latter is dominated by the operators
$(\partial_{i}\theta\partial_{i}\theta)^{n}$; see (\ref{HZ1}).
The cancellation
becomes possible due to the fact that the dimension $\Delta_{n}$
for $\alpha=0$ is a function linear in $n$, see (\ref{simple}),
and therefore all the terms in the square brackets in (\ref{struc2})
are independent of $\Lambda r$. In this case,
the anomalous exponents are determined by the critical dimensions
of the powers of the operator $\partial_{i}\theta\partial_{i}\theta$;
these dimensions are known up to the order $\eps^{2}$ of the
$\eps$ expansion \cite{RG}.

For $d=1$, the behavior analogous to (\ref{HZ1}) in the model
(\ref{4}) is demonstrated by the structure functions of the
field $\phi(t,x)$ defined so that  $\theta (t,x)=\partial_{x}
\phi(t,x)$. In this formulation, the problem was studied in
the paper \cite{VM} within the zero-modes approach and using
numerical simulations. The structure functions of the ``potential''
$\phi$ are not simply related to the structure functions of the
primary field $\theta$, but they can be derived directly using the
RG technique. Obviously, the field $\phi$ enters into the vertex
in the form of the derivative
$\theta'\partial_{x}( v\partial_{x}\phi)$. Therefore, the
operators $\phi^{n}$ are not renormalized and their critical
dimensions are given by the relations analogous to (\ref{simple}):
$\Delta[\phi^{n}]=n\Delta[\phi]$, where $\Delta[\phi] =-1+
\Delta[\theta]=-2+\eps/2$, see Eq. (\ref{33b}). The structure
functions are then given by the expression analogous to (\ref{4}):
\[ S_{2n}\simeq D_{0}^{-n} r^{n(4-\eps)} (Mr)^{\Delta_{2n}},  \]
where the part of the anomalous exponents is played by the critical
dimensions $\Delta_{2n}$ of the operators
$(\partial_{x} \phi \partial_{x} \phi )^{n}\equiv \theta^{2n}$
given by Eq. (\ref{d1}). In the notation of \cite{VM} we then
have $\zeta_{2n}=n(4-\eps)+ \Delta_{2n}=2n-n\eps(2n-1) -
2\pi n(n-1)(2n-1)\eps^{2}/3\sqrt{3}$, in agreement with the
$O(\eps)$ result obtained in \cite{VM} using the zero-modes approach;
the exponent $\zeta_{2}=2-\eps$ is exact.

Let us conclude this section with a brief discussion of the simple
model of a passively advected magnetic field considered
in \cite{MHD}.\footnote{
In more realistic models of the MHD turbulence the magnetic field
indeed behaves as a passive vector in the
so-called kinetic fixed point of the RG equations, see
\cite{MHD1,MHD2}.}
In this case, both $\theta\equiv\theta_{i}(x)$ and
the velocity are solenoidal vector fields.
The velocity field
is taken to be Gaussian with the correlator
(\ref{3}), and the nonlinearity in  (\ref{1}) has the form
$v_j\partial_{j}\theta_{i}-\theta_{j}\partial_{j}v_i$.
The anomalous scaling in this model also appears already
for the pair correlator; the corresponding exponent is found exactly
\cite{MHD}.

The RG analysis given above and in \cite{RG} is extended directly
to this model. It turns out that the expressions for
the renormalization constant $Z_{\nu}$,
the RG functions  $\beta$ and $\gamma_{\nu}$,
the fixed point $g_{*}$, and the critical dimension $\Delta_{\theta}$
coincide exactly with the expressions (\ref{RGF})--(\ref{fixed})
and (\ref{33b})
for the model (\ref{action})  with the substitution $\alpha=0$.
For the IR asymptotic region, the expressions of
the form (\ref{55})  are obtained for the correlation functions of various
composite operators; the corresponding critical dimensions $\Delta_{F}$
are calculated in the form of the $\eps$ expansions. In particular,
for the critical dimensions $\Delta_{2n}$ of the scalar operators
$\theta^{2n}\equiv(\theta_{i}\theta_{i})^{n}$  we have obtained
\begin{equation}
\Delta_{2n}= -2n-\frac{2n(n-1)\eps}{d+2}+O(\eps^{2}),
\label{figvam1}
\end{equation}
and for the special case $n=1$ we have
\begin{equation}
\Delta_{2}= -2-\frac{2(d-2)\eps^{2}}{d(d-1)}+O(\eps^{3}).
\label{figvam11}
\end{equation}
For the dimensions $\Delta_{2n}'$  of the second-rank irreducible
tensors
$\theta_{i}\theta_{j} \theta^{2n-2}-\delta_{ij}\theta^{2n}/d$ we have
\begin{equation}
\Delta_{2n}'= -2n+\frac{\eps[d(d+1)-2(d-1)n(n-1)]}
{(d-1)(d+2)}+O(\eps^{2}).
\label{figvam2}
\end{equation}
The leading terms of the small $Mr$  behavior of the scaling functions
are determined by the contributions of the scalar operators $\theta^{2n}$,
and the part of the anomalous exponents is played by the dimensions
(\ref{figvam1}). For the special case of the pair correlator it then follows
$\langle\theta(x)\theta(x')\rangle \propto (\Lambda r)^{-2\Delta_{\theta}}
(Mr)^{\Delta_{2}}$. In the notation of \cite{MHD} we have
$\gamma=\Delta_{2}-2\Delta_{\theta}$; from Eqs. (\ref{33b}) and
(\ref{figvam11}) it follows
$\Delta_{2}-2\Delta_{\theta}=-\eps-2\eps^{2}(d-2)/
{d(d-1)}+O(\eps^{3})$ for any $d$,
in agreement with the exact expression for  $\gamma$
obtained in \cite{MHD}.

\section{Calculation of the anomalous exponents to the order $\eps^{2}$}
\label {sec:5}

In this section we present the two-loop calculation of the critical
dimensions $\Delta_{n}$ of the composite operators  $\theta^{n}$,
which determine the anomalous exponents in the expressions
(\ref{as1}), (\ref{as2}).

The operators $\theta^{n}$ are renormalized multiplicatively,
$\theta^{n}=Z_{n}[\theta^{n}]_{ren}$
(see Sec. \ref{sec:3}). The renormalization constants $Z_{n}$
can be found from the requirement that the 1-irreducible
correlation function
\begin{equation}
\langle [\theta^{n}]_{ren}(x) \theta(x_{1})\dots\theta(x_{2})
\rangle_{1-ir}=  Z_{n}^{-1}\langle \theta^{n}(x)
\theta(x_{1})\dots\theta(x_{2})\rangle_{1-ir}
\equiv Z_{n}^{-1}\Gamma_{n}
\label{req}
\end{equation}
be UV finite, i.e., have no poles in  $\eps$,
when expressed in renormalized variables using the formulas
(\ref{reno}).
This requirement determines  $Z_{n}$
up to an UV finite part; the choice of the finite part
depends on the ``subtraction scheme''. Most convenient for
practical calculations is the minimal subtraction (MS) scheme.
In the MS scheme, only poles in  $\eps$
are subtracted from the divergent expressions, and the
renormalization constants have the form ``1 + only poles in $\eps$''.
In particular,
\begin{eqnarray}
Z_{n}^{-1}=1+\sum _{k=1}^{\infty }a_k(g)\eps^{-k}=
1+\sum _{n=1}^{\infty }g^n
\sum _{k=1}^{n}a_{nk}\eps ^{-k}.
\label{1.30}
\end{eqnarray}
The coefficients $a_{nk}$ in our model depend only on the space
dimension $d$ and the completely dimensionless parameter $\alpha$;
their independence of $\eps$ is a feature specific for the
MS scheme.  One-loop diagrams generate contributions of order
$g$ in (\ref{1.30}), two-loop ones generate contributions of order
$g^2$, and so on. The order of the pole in $\eps$ does not exceed
the number of the loops in the diagram.

The two-loop diagrams of the function $\Gamma_{n}$ required
for the calculation of $Z_{n}$ to the order $g^2$
and the corresponding symmetry coefficients are given in Table II.
The solid lines in the diagrams denote the bare propagator
$\langle\theta\theta'\rangle_{0}$ from (\ref{lines1}), the end
with a slash correspond
to the field $\theta'$, and the end without a slash correspond
to $\theta$; the dashed lines denote the bare propagator
(\ref{4}). Note that the propagator $\langle\theta\theta\rangle_{0}$
does not enter into the diagrams for $\Gamma_{n}$.
The black circle with $p\ge0$ attached ``legs'' denotes the vertex
factor $F_{p}$ given by the $p$-fold variational derivative
$ F_{p}\equiv \delta \theta^{n}(x)/\delta \theta(x_{1})
\dots \delta \theta(x_{p})$.

Now let us turn to the calculation of the diagrams from Table II.
It is sufficient to calculate the function $\Gamma_{n}$ in the
momentum-frequency representation with all the external momenta
and frequencies equal to zero; the IR regularization is then
provided by the ``mass'' $m$ from the correlator (\ref{4}).
In what follows, we use the notation
\begin{mathletters}
\label{not}
\begin{equation}
R_{ij}({\bf k})\equiv D_{0} P_{ij} ({\bf k})+D_{0}' Q_{ij} ({\bf k})
\label{not1}
\end{equation}
and
\begin{equation}
S(k)\equiv (k^{2}+m^{2})^{-d/2-\eps/2} .
\label{not2}
\end{equation}
\end{mathletters}
We also recall the relations
$D_{0}=g_{0} \nu_0$ and $\alpha=D_{0}'/ D_{0}$.

The diagram $D_{2}$ differs from $D_{1}$ only by the insertion
of the simplest self-energy diagram $\Sigma_{\theta\theta'}$
into one of the two lines $\langle\theta\theta'\rangle$.
Therefore, the combination $D_{1}+2D_{2}$ entering into $\Gamma_{n}$
can be easily calculated as a whole: we calculate the
single-loop diagram $D_{1}$  with the {\it exact} propagators
$\langle\theta\theta'\rangle$
instead of the bare propagators $\langle\theta\theta'\rangle_{0}$
and then expand the result in $g_{0}$
to the order $g_{0}^{2}$.  From the exact solution (see Sec. \ref{sec:2})
it follows that the propagator $\langle\theta\theta'\rangle$
is obtained from its bare counterpart simply by the replacement
$\nu_0\to\eta_{0}$, where the exact ``effective diffusivity''
has the form\footnote{ See \cite{BC} for the exact expression
for the effective diffusivity in the incompressible case.}
\[\eta_{0}\equiv\nu_0 + \frac{D_{0}(d-1)+D_{0}'}{2d} J(m),\]
see (\ref{Dyson1}), (\ref{otvet}). Then the ``exact''  analogue
of the diagram $D_{1}$ is given by
\begin{equation}
\int\frac{d\omega}{2\pi}\int\frac{d{\bf k}}{(2\pi)^{d}}\,
 \frac{S(k)\, R_{ij}\, ({\bf k}) k_{i}k_{j}}
{|{\rm i}\omega+\eta_{0}k^{2}|^{2}}
= D_{0}' J(m) /2\eta_{0},
\label{int1}
\end{equation}
where we have performed the elementary integration over the frequency
and used the isotropy of the function $S(k)$.
The expansion of the result (\ref{int1}) in $g_{0}$ gives
\begin{equation}
D_{1}+2D_{2} = \frac{\alpha g_{0} J(m)}{2} \left[1-\frac{g_{0}
(d-1+\alpha)J(m)} {2d} \right].
\label{cont1}
\end{equation}
The right hand side of Eq. (\ref{cont1}) is expressed
in renormalized variables
by the substitution $g_{0}=g\mu^\eps Z_{\nu}^{-1}$ with the constant
$Z_{\nu}$ from (\ref{Z}), which within our accuracy gives:
\begin{eqnarray}
D_{1}+2D_{2} = \frac{\alpha g\mu^\eps J(m)}{2}  +
\frac{\alpha\,g^{2}(d-1+\alpha)\mu^\eps J(m)} {4d}
\left[C_{d}/\eps -\mu^\eps J(m)
 \right] \equiv
\nonumber \\
\equiv  g\, D^{(1)} + g^{2}\, D^{(2)}.
\label{cont1R}
\end{eqnarray}

The diagram $D_{3}$ is represented by the integral
\begin{equation}
D_{3}= \int\frac{d\omega}{2\pi}\int\frac{d\omega'}{2\pi}
\int\frac{d{\bf k}}{(2\pi)^{d}}\int\frac{d{\bf q}}{(2\pi)^{d}}\,
\frac{R_{ij}({\bf q}) (k+q)_{i}(k+q)_{j} R_{ps}({\bf k})k_{p}k_{s}}
{|{\rm i}\omega+\nu_{0}({\bf k}+{\bf q})^{2}|^{2}\,
|{\rm i}\omega'+\nu_{0}k^{2}|^{2}}  S(k) S(q),
\label{int2}
\end{equation}
and the integrations over the frequencies give
\begin{equation}
D_{3}=  \frac{\alpha g_{0}^{2}}{4}
\int\frac{d{\bf k}}{(2\pi)^{d}}\int\frac{d{\bf q}}{(2\pi)^{d}}\,
\left[\alpha + (1-\alpha)\frac{k^{2}\sin^{2}\vartheta}
{({\bf k}+{\bf q})^{2}} \right]S(k) S(q),
\label{int3}
\end{equation}
where $\vartheta$ is the angle between the vectors ${\bf k}$ and
${\bf q}$, so that ${\bf k}\cdot{\bf q}=kq\cos\vartheta$.
The symmetry of the integral (\ref{int3}) in ${\bf k}$ and ${\bf q}$
allows one to perform the substitution
$k^{2}\to  {({\bf k}+{\bf q})^{2}}/2- {\bf k}\cdot{\bf q}$
in the integrand, which gives:
\begin{equation}
D_{3}=  \frac{\alpha^{2} g_{0}^{2}}{4} J^{2}(m)+
\frac{\alpha (1-\alpha) g_{0}^{2}}{8} [J_{1}(m)-2J_{2}(m)],
\label{int4}
\end{equation}
where we have written
\begin{equation}
J_{1}(m)\equiv
\int\frac{d{\bf k}}{(2\pi)^{d}}\int\frac{d{\bf q}}{(2\pi)^{d}}\,
\sin^{2}\vartheta\,S(k) S(q)
\label{int5}
\end{equation}
and
\begin{equation}
J_{2}(m)\equiv
\int\frac{d{\bf k}}{(2\pi)^{d}}\int\frac{d{\bf q}}{(2\pi)^{d}}\,
\frac{{\bf k}\cdot{\bf q}\, \sin^{2}\vartheta}
{({\bf k}+{\bf q})^{2}} \,S(k) S(q).
\label{int6}
\end{equation}
The integral in (\ref{int5}) can be easily expressed via $J(m)$:
\begin{eqnarray}
J_{1}(m)=C^{2}_{d} \int_{0}^{\infty} dk\, k^{d-1}
\int_{0}^{\infty} dq\, q^{d-1}  \int d{\bf n}
\sin^{2}\vartheta\,S(k) S(q)=
\nonumber \\
=J^{2}(m) \int d{\bf n}\sin^{2}\vartheta=
\frac{d-1}{d}J^{2}(m),
\label{int7}
\end{eqnarray}
with the coefficient $C_{d}$ from  (\ref{Z}). Here and below
$\int d{\bf n}$ denotes the integral over the $d$-dimensional
sphere, normalized with respect to its area, so that
$\int d{\bf n}\,1=1$ and $\int d{\bf n}\sin^{2}\vartheta=
(d-1)/d$.
For the integral (\ref{int6})  one has:
\begin{eqnarray}
J_{2}(m)=C^{2}_{d} \int_{0}^{\infty} dk \int_{0}^{\infty} dq  \int d{\bf n}
\frac {k^{d}q^{d}\,\cos\vartheta\,\sin^{2}\vartheta}
{k^{2}+q^{2}+2kq\cos\vartheta}
\,S(k) S(q)=
\nonumber \\
= 2 C^{2}_{d} \int_{0}^{\infty} dk \int_{0}^{k} dq \int d{\bf n}
\frac {k^{d}q^{d}\,\cos\vartheta\,\sin^{2}\vartheta}
{k^{2}+q^{2}+2kq\cos\vartheta}
\,S(k) S(q),
\label{int8}
\end{eqnarray}
where we have used the symmetry of the integrand and
integration area in $k$ and $q$.

In order to find the renormalization constant,
we need not the entire exact
expression (\ref{int8}) for the integral $J_{2}(m)$, we rather
need its UV divergent part. The simple power counting shows that
the UV divergence of the integral (\ref{int8}) is generated by
the region in which the both integration momenta
$k$ and $q$ are large. Therefore, the integral (\ref{int8})
contains only a first-order pole in $\eps$, and the coefficient in
$1/\eps$ does not change when the integration area $[0,\infty]$
for the momentum $k$ is restricted from below by some finite limit,
for example, $[m,\infty]$. Furthermore, the IR regularization of
the integral is then provided by this finite lower limit, and
one can simply set $m=0$ in the functions $S(k)$, $S(q)$, which gives:
\begin{equation}
J_{2}(m)\simeq 2 C^{2}_{d} \int_{m}^{\infty} dk \int_{0}^{k} dq
\int d{\bf n} \,
\frac{k^{-\eps}q^{-\eps}\,\cos\vartheta\,\sin^{2}\vartheta}
{k^{2}+q^{2}+2kq\cos\vartheta}.
\label{int9}
\end{equation}
Here and below $\simeq$ means the equality up to the terms
finite for $\eps\to0$.
From the dimensionality considerations, it is obvious that
$J_{2}(m)=m^{-2\eps} f(\eps)$, where $f(\eps)$ contains a first-order
pole in $\eps$. It then follows that
\begin{equation}
J_{2}(m)=-\frac{1}{2\eps} {\cal D}_{m}J_{2}(m)
\label{int10}
\end{equation}
(we recall the notation ${\cal D}_{m}\equiv m\partial/\partial m$).
The representation (\ref{int10}) allows one to get rid of the integration
over $k$ in (\ref{int9}):
\begin{equation}
J_{2}(m)\simeq C^{2}_{d}\,\frac{m^{-2\eps}}{\eps} \int_{0}^{1} dx\,
\int d{\bf n} \,\frac{x^{-\eps}\,\cos\vartheta\,\sin^{2}\vartheta}
{1+x^{2}+2x\cos\vartheta},
\label{int11}
\end{equation}
where we have performed the substitution $q\equiv mx$.
The pole in (\ref{int11}) is isolated explicitly, the integral is
UV convergent and one can set $\eps=0$ in the integrand:
\begin{equation}
J_{2}(m)\simeq C^{2}_{d}\,\frac{m^{-2\eps}}{\eps} \int_{0}^{1} dx\,
\int d{\bf n}\,
\frac{\cos\vartheta\,\sin^{2}\vartheta}
{1+x^{2}+2x\cos\vartheta}.
\label{int12}
\end{equation}
The integrations in (\ref{int12}) are performed explicitly:
\begin{equation}
J_{2}(m)\simeq C^{2}_{d}\,\frac{m^{-2\eps}}{2\eps} \int d{\bf n}\,
\vartheta\,\cos\vartheta\,\sin\vartheta= C^{2}_{d}\,
\frac{m^{-2\eps}(1-d)}{2\eps d^{2}}.
\label{int13}
\end{equation}
Combining the expressions (\ref{int4}), (\ref{int7}), and (\ref{int13})
we obtain:
\begin{equation}
D_{3}=J^{2}(m)g_{0}^{2} \left[\frac{\alpha^{2}}{4}
+\frac{\alpha(1-\alpha)(d-1)} {8d} \right] + g_{0}^{2} C_{d}^{2}\,
\frac{\alpha(1-\alpha)(d-1)\,m^{-2\eps}}{8\eps d^{2}}.
\label{int14}
\end{equation}
Within our accuracy, the renormalization of the expression (\ref{int14})
is reduced to the substitution $g_{0}\to g \mu^{\eps}$, which
gives:
\begin{equation}
D_{3}=J^{2}(m)\mu^{2\eps} \left[\frac{\alpha^{2}g^{2}}{4}+\frac{\alpha g^{2}
(1-\alpha)(d-1)} {8d} \right] + g^{2}C_{d}^{2}\,
\frac{\alpha(1-\alpha)(d-1)(\mu/m)^{2\eps}}{8\eps d^{2}}.
\label{int15}
\end{equation}

Now let us turn to the diagram $D_{4}$. It is given by the expression
\begin{eqnarray}
D_{4} =\int\frac{d\omega}{2\pi}\int\frac{d\omega'}{2\pi}
\int\frac{d{\bf k}}{(2\pi)^{d}}\int\frac{d{\bf q}}{(2\pi)^{d}}\times
\nonumber \\
\times\frac{
R_{ij}({\bf k}) k_{i}(k+q)_{j}
R_{ps}({\bf k})q_{p}q_{s} S(k) S(q)} {
({\rm i}\omega+\nu_{0}k^{2}) |{\rm i}\omega'+\nu_{0}q^{2}|^{2}
(-{\rm i}(\omega+\omega')+\nu_{0}({\bf k}+{\bf q})^{2})}=
\nonumber \\
=\frac{\alpha^{2}g_{0}^{2}}{8} [J^{2}(m)+J_{3}(m)],
\label{int16}
\end{eqnarray}
where we have performed the integrations over the frequencies and
made use of the symmetry in $k$ and $q$; the integral $J_{3}(m)$
is given by:
\begin{equation}
J_{3}(m)\equiv
\int\frac{d{\bf k}}{(2\pi)^{d}}\int\frac{d{\bf q}}{(2\pi)^{d}}\,
\frac{ {\bf k}\cdot{\bf q}\,S(k) S(q)} {k^{2}+q^{2}+{\bf k}\cdot{\bf q}}.
\label{int17}
\end{equation}
Proceeding as for the integral $J_{2}(m)$  above,
we arrive at the expression
\begin{equation}
J_{3}(m)\simeq C_{d}^{2}\,
\frac{m^{-2\eps}}{\eps}  \int_{0}^{1} dx\, \int d{\bf n}\,
\frac{\cos\vartheta} {1+x^{2}+x\cos\vartheta},
\label{int18}
\end{equation}
which is analogous to the expression (\ref{int12}) for $J_{2}(m)$.
In contrast to (\ref{int12}), after the integration over $x$ in
(\ref{int18})  we arrive at the
integral over the angles which cannot be calculated explicitly.
We rather expand the integrand in (\ref{int18}) in $\cos\vartheta$:
\begin{eqnarray}
\int_{0}^{1} dx\, \int d{\bf n}\,
\frac{\cos\vartheta} {1+x^{2}+x\cos\vartheta}=
\nonumber \\
=\int_{0}^{1} dx\, \int d{\bf n}\,  \frac{\cos\vartheta}{1+x^{2}} \,
\sum_{k=0}^{\infty} \left(-\frac{x\,\cos\vartheta}{1+x^{2}}
\right)^{k},
\label{int19}
\end{eqnarray}
and use the formulas
\begin{eqnarray}
\int d{\bf n}\,\cos^{2k}\vartheta=\frac{(2k-1)!!}{d(d+2)\dots (d+2k-2)},
\nonumber \\
\int d{\bf n}\,\cos^{2k+1}\vartheta= 0,
\nonumber \\
\int_{0}^{1} dx\, \frac{x^{2k+1}}{(1+x^{2})^{2k+2}}=
\frac{(k!)^{2}} {4 (2k+1)!}.
\label{int20}
\end{eqnarray}
For the series in (\ref{int19}) this gives
(we omit an overall minus sign):
\begin{eqnarray}
\frac{1}{4d}\sum_{k=0}^{\infty}\frac{(2k+1)!! (k!)^{2}}
{(2k+1)! (d+2)\dots(d+2k)}=
\frac{1}{4d}\sum_{k=0}^{\infty} \frac{k!}{4^{k} (d/2+1)\dots(d/2+k)}
\equiv h(d)/4d,
\label{int22}
\end{eqnarray}
with the function $h(d)$  entering into the expressions
(\ref{Dn})--(\ref{Gamma1}).

Combining the expressions (\ref{int16}), (\ref{int18}), (\ref{int22}),
and performing the replacement $g_{0}\to g \mu^{\eps}$
we obtain
\begin{equation}
D_{4}=J^{2}(m)\mu^{2\eps}\frac{\alpha^{2}g^{2}}{8} -
\frac{\alpha^{2}g^{2}h(d)C_{d}^{2}(\mu/m)^{2\eps}}{32d\eps}.
\label{int23}
\end{equation}

The diagram $D_{5}$ is simply given by
\begin{equation}
D_{5}=D_{1}^{2}=  J^{2}(m)\mu^{2\eps}\frac{\alpha^{2}g^{2}}{4},
\label{int24}
\end{equation}
and $D_{6}$ contains effectively a closed circuit of retarded
propagators and vanishes identically. Therefore, the function
$\Gamma_{n}$ in the two-loop order of the renormalized perturbation
theory has the form
\begin{equation}
\Gamma_{n} = 1+ \frac{n(n-1)}{2} (D_{1}+2D_{2}+D_{3})+
n(n-1)(n-2)D_{4}+\frac{n(n-1)(n-2)(n-3)}{8} D_{5},
\label{rengam}
\end{equation}
with the symmetry coefficients from Table II and the explicit expressions
for $D_{i}$ given in (\ref{cont1R}), (\ref{int15}), (\ref{int23}),
and (\ref{int24}).

Within our accuracy, the renormalization constant (\ref{1.30})
has the form
\begin{equation}
Z^{-1}_{n}=1+\frac{a_{11}g}{\eps}+\frac{a_{21}g^{2}}{\eps} +
\frac{a_{22}g^{2}}{\eps^{2}} +O(g^{3}),
\label{ZF}
\end{equation}
and the requirement that the function (\ref{req}) be UV finite
in the first order in $g$  gives:
\begin{equation}
\frac{a_{11}g}{\eps}+\frac{n(n-1)}{2} \,g D^{(1)}
= {\ \rm UV\ finite},
\label{fini1}
\end{equation}
with the coefficient $D^{(1)}$ defined in (\ref{cont1R}).
The expansion in $\eps$ of the integral $J(m)$ from (\ref{otvet})
entering into the expressions for $D_{i}$ has the form
\begin{equation}
\mu^{\eps} J(m)=\frac{C_{d}}{\eps}\left[1+
\eps\left(\frac{\psi(1)-\psi(d/2)}{2}+
\ln(\mu/m) \right)\right] +O(\eps),
\label{Jexp}
\end{equation}
where $\psi(z)\equiv d\ln \Gamma(z)/ dz$.
From (\ref{cont1R}), (\ref{fini1}), and the first term of the expansion
(\ref{Jexp}) one obtains:
\begin{equation}
a_{11}=-\alpha n(n-1)C_{d}/4.
\label{a1}
\end{equation}
The UV finiteness of the function (\ref{req}) in the
order  $g^{2}$ implies:
\begin{eqnarray}
\frac{a_{21}g^{2}}{\eps} +\frac{a_{22}g^{2}}{\eps^{2}}+\frac{a_{11}g}{\eps}
\ \frac{n(n-1)}{2}\, g D^{(1)}
+\frac{n(n-1)}{2} (g^{2}\,D^{(2)}+D_{3})+
\nonumber \\
+n(n-1)(n-2)D_{4}+\frac{n(n-1)(n-2)(n-3)}{8} D_{5}= {\ \rm UV\ finite},
\label{fini2}
\end{eqnarray}
which along with the expressions (\ref{cont1R}), (\ref{int15}),
(\ref{int23}),
(\ref{int24}), (\ref{a1}) and the expansion (\ref{Jexp}) yields:
\begin{mathletters}
\label{a}
\begin{equation}
a_{21}/C_{d}^{2}= \frac{n(n-1)\alpha(\alpha-1)(d-1)}{16d^{2}} +
\frac{n(n-1)(n-2)\alpha^{2}\,h(d)}{32d},
\label{a21}
\end{equation}
\begin{eqnarray}
a_{22}/C_{d}^{2}= \frac{\alpha^{2}n^{2}(n-1)^{2}}{16}-
\frac{\alpha^{2}n(n-1)}{8}+\frac{n(n-1)\alpha(\alpha-1)(d-1)}{16d} -
\nonumber \\
-\frac{\alpha^{2}n(n-1)(n-2)}{8}-\frac{\alpha^{2}n(n-1)(n-2)(n-3)}{32}=
\nonumber \\
= \frac{\alpha\,n(n-1)}{32} [\alpha\,n(n-1) -
2(\alpha+d-1)/d \,].
\label{a22}
\end{eqnarray}
\end{mathletters}
We note that the $O(1)$ terms of the expansion (\ref{Jexp})
cancel out in the Eq. (\ref{fini2}) and therefore give no
contribution to the coefficients $a_{ij}$.

For the corresponding anomalous dimension
$\gamma_{n}\equiv\widetilde{\cal D}_{\mu}\ln Z_{n}$
we have:
\begin{equation}
\gamma_{n}\equiv\widetilde{\cal D}_{\mu}\ln Z_{n}=\beta(g)\partial_{g}
\ln Z_{n}=  [-\eps+\gamma_{\nu}(g)] {\cal D}_{g}\ln Z_{n},
\label{canc}
\end{equation}
with the RG functions $\beta(g)$ and $\gamma_{\nu}(g)$ from (\ref{RGF}).
Within our accuracy (\ref{canc}) yields
\begin{equation}
\gamma_{n}=a_{11}g+2a_{21}g^{2} +\frac{1}{\eps}\left[
-a_{11}g\gamma_{\nu}+ 2a_{22}g^{2}-(a_{11}g)^{2} \right],
\label{canc2}
\end{equation}
and using the explicit expressions (\ref{a1}), (\ref{a})
one obtains:
\begin{equation}
\gamma_{n}=\frac{-\alpha n(n-1)u}{4}+
\frac{n(n-1)\alpha(\alpha-1)(d-1)u^{2}}
{8d^{2}}+ \frac{n(n-1)(n-2)\alpha^{2}h(d)u^{2}}{16d}+O(g^{3}),
\label{otvetgamma}
\end{equation}
where $u\equiv g C_{d}$.
It follows from the explicit expressions
(\ref{gammanu}), (\ref{a1}), and (\ref{a})  that the
coefficient in $1/\eps$ in the expression (\ref{canc2})
for $\gamma_{n}$  vanishes:
$-a_{11}g\gamma_{\nu}+ 2a_{22}g^{2}-(a_{11}g)^{2}=0$.
This is a manifestation of the general fact that the function
$\gamma_{n}$ is UV finite, i.e., it has no poles in $\eps$.

Substituting the anomalous dimension (\ref{otvetgamma}) into
the expression (\ref{32B}) and performing the replacement
$g\to g_{*}$ with $g_{*}$ from (\ref{fixed}), we arrive at
the desired expression (\ref{Dn}) for the critical dimension
of the composite operator $\theta^{n}$.

It is worth noting that the case $d=1$ is exceptional in the
sense that ``there are no angles in one dimension.''  We have
performed all the calculation directly in $d=1$ and checked that
the one-dimensional exponents are indeed obtained from the
general expressions like (\ref{dip}) by the substitution $d=1$.

\section{Discussion and conclusion}
\label {sec:6}

We have applied the RG and OPE methods to the simple model
(\ref{1}), (\ref{2}), (\ref{4}), which describes the advection
of a passive scalar by the non-solenoidal (``compressible'')
velocity field, decorrelated in time and self-similar in space.
We have shown that the correlation functions of the scalar field
in the convective range exhibit anomalous scaling behavior;
the corresponding anomalous exponents have been calculated
to the second order of the $\eps$ expansion
(the two-loop approximation), see (\ref{Dn})--(\ref{d1}).
They depend on a free parameter, the ratio $\alpha=D_{0}'/D_{0}$
of the amplitudes in the transversal and longitudinal parts
of the velocity correlator, and are in this sense non-universal.
In the language of the RG, the non-universality of the exponents
is related to the fact that the fixed point of the RG equations
is degenerated: its coordinate depends continuously on $\alpha$.

In contrast to the model (\ref{3}), where the anomalous exponents
are determined by the critical dimensions of the composite
operators $(\partial_{i}\theta\partial_{i}\theta)^{n}$,
the exponents in the model (\ref{4}) are related to the critical
dimensions  of the monomials $\theta^{n}$, the powers of the field
itself, and these dimensions appear to be nonlinear functions of
$n$. This explains the important difference between the anomalous
scaling behavior of the model (\ref{3}) and that of the model
(\ref{4}): in the latter, the correlation functions in the
convective range depend substantially on both the IR and UV
characteristic scales, and the structure functions are independent
of the separation $r=|{\bf x}-{\bf x'}|$. The monomials $\theta^{n}$
in the model (\ref{4}) also provide an example of the power
field operators {\it without derivatives}, whose correlation
functions exhibit multifractal behavior (another interesting
example is the field theoretical model of a growth process
considered in \cite{Hol}). Analogous behavior is demonstrated
by the model of a magnetic field, advected passively by the
incompressible Gaussian velocity; the corresponding anomalous
exponents are calculated to the order $\eps$ ($\eps^{2}$ for
the pair correlator).

The anomalous exponent for the pair correlation function has
been found exactly for all $0<\eps<2$. Its expansion in $\eps$
coincides with the result obtained using the RG for all values
of the space dimensionality $d$ and ratio $\alpha$. The agreement
between the exact exponent for the pair correlation function and
the first two terms of the corresponding $\eps$ expansion is also
established for a passively advected magnetic field.

These facts support strongly the applicability of the RG technique
and the $\eps$ expansion to the problem of anomalous scaling for
the finite values of $\eps$, at least for low-order correlation
functions.

We note that the series in $\eps$ for all known exact exponents
in the rapid-change models have finite radii of convergence, a rare
thing for field theoretical models. In the language of the field theory,
this is related to the fact that in the rapid-change models, there is
no factorial growth of the number of diagrams in higher orders
of the perturbation theory (a great deal of diagrams indeed vanish
owing to retardation, see discussion in Sec. \ref{sec:2}). In its turn,
this fact suggests that the series in $\eps$ for the unknown exponents
(for example, the anomalous exponents in the original
Obukhov--Kraichnan model) can also be convergent.

It should also be noted that the asymptotic expressions
(\ref{as1}), (\ref{as2}) result from the fact that the critical
dimensions $\Delta_{n}$ are negative, and that the modulus
$|\Delta_{n}|$ increases monotonously with $n$. This is obviously
so within the $\eps$ expansion, in which the sign and the $n$
dependence of the dimensions are determined by the first-order
terms (\ref{exp}), (\ref{figvam1}), while the higher-order terms
are treated as small corrections. However, for finite values of
$\eps$ the higher-order terms can, in principle, change these
features of the dimensions. Indeed, the $n^{3}$ contribution
in the second-order approximation for $\Delta_{n}$ is positive,
see e.g. (\ref{Dn}), so that $\Delta_{n}$ also becomes positive,
provided $n$ is large enough. Of course, this conclusion is
based on the second-order approximation of the $\eps$  expansion
and is therefore not definitive: the higher-order terms of the $\eps$
expansion contain additional powers of $n$, so that the actual
expansion parameter appears to be $\eps n$ rather than
$\eps$ itself, cf.\cite{RG,Hol}.
Therefore, the correct analysis of the large $n$
behavior of the anomalous exponents requires resummation of the
$\eps$ expansions with the additional condition that $\eps n\simeq1$.
This is clearly not a simple problem and it requires considerable
improvement of the existing technique.

\acknowledgments
The authors are thankful to A.~N.~Vasil'ev for clarifying discussions
and to A.~V.~Runov for his help in preparing the Figures.
One of the authors (N.V.A.)  is thankful to G.~L.~Eyink and
R.~H.~Kraichnan for interesting remarks regarding the paper \cite {RG}.

The work was supported by the Russian Foundation for Fundamental
Research (grant N  96-02-17-033) and by the Grant Center for
Natural Sciences of the Russian State Committee for Higher Education
(grant N 97-0-14.1-30).

\begin{table}
\caption{Canonical dimensions of the fields and parameters in the
model (2.1).}
\label{table1}
\begin{tabular}{ccccccccc}
$F$ & $\theta $ & $\theta '$ & $ {\bf v} $ &
$\nu ,\nu _{0}$ &$D_{0},D_{0}'$
& $m,M,\mu,\Lambda$ & $g_{0}$ & $g,\alpha$ \\
\tableline
$d_{F}^{k}$       & 0    & d   & -1 & -2 &-2+$\eps$  & 1 & $\eps$ & 0 \\
$d_{F}^{\omega }$ & -1/2 & 1/2 & 1  & 1  & 1 & 0 & 0 & 0 \\
$d_{F}$           & -1   & d+1 & 1  & 0  & $\eps$  & 1 & $\eps$ & 0 \\
\end{tabular}
\end{table}

\vskip1cm

\begin{table}
\caption{The diagrams of the 1-irreducible Green function
$\langle \theta^{n}(x) \theta(x_{1})\dots\theta(x_{2})\rangle$
in the two-loop approximation.}
\label{table2}
\begin{tabular}{ccc}
{} & Diagram & Symmetry coefficient \\
\tableline \\
$D_{1}$ &  \dA & $n(n-1)/2$ \\
$D_{2}$ &  \dC & $n(n-1)$ \\
$D_{3}$ &  \dB & $n(n-1)/2$ \\
$D_{4}$ &  \dE & $n(n-1)(n-2)$ \\
$D_{5}$ &  \dF & $n(n-1)(n-2)(n-3)/8$ \\
$D_{6}$ &  \dD & --- \\
\end{tabular}
\end{table}
\end{document}